\documentclass[fleqn,usenatbib]{mnras}

\usepackage{newtxtext,newtxmath}
\usepackage{multirow}

\usepackage[T1]{fontenc}
\usepackage{ae,aecompl}

\usepackage{graphicx}	% Including figure files
\usepackage{amsmath}	% Advanced maths commands

\usepackage{amssymb}	% Extra maths symbols

\newcommand{\hii}    {H\,{\sc{ii}}}
\newcommand{\mkm}    {$\mu$m}
\newcommand{\tgas}   {$T_{\rm gas}$}

\newcommand{\kms}    {km\,s$^{-1}$}
\newcommand{\co}     {$^{13}$CO}
\newcommand{\cvo}    {C$^{18}$O}
\newcommand{\cs}     {C$^{34}$S}

\newcommand{\meth}   {CH$_3$OH}

\usepackage{xcolor}
\usepackage{ulem}

\title[The warm-up phase in cores around RCW\,120]{The warm-up phase in massive star-forming cores around RCW\,120}

\author[M. S. Kirsanova et al.]{
M. S. Kirsanova,$^{1,2}$\thanks{E-mail: kirsanova@inasan.ru}
S. V. Salii$^{3}$,
S. V. Kalenskii$^{4}$, D. S. Wiebe$^{1}$,
A. M. Sobolev$^{3}$,
P. A. Boley$^{2,3}$
\\
$^{1}$Institute of Astronomy, Russian Academy of Sciences, 119017, 48 Pyatnitskaya Str., Moscow, Russia\\
$^{2}$Moscow Institute of Physics and Technology, 141701, 9 Institutskiy per., Dolgoprudny, Moscow Region, Russia\\
$^{3}$Institute of Natural Sciences and Mathematics, Ural Federal University,
19 Mira Str., 620075 Ekaterinburg, Russia\\
$^{4}$Astro Space Center, Lebedev Physical Institute, Russian Academy of Sciences, 117997, 84/32 Profsoyuznaya Str., Moscow, Russia\\
}

\date{Accepted \today. Received \today; in original form \today}

\pubyear{2020}

\begin{document}
\label{firstpage}
\pagerange{\pageref{firstpage}--\pageref{lastpage}}
\maketitle

\begin{abstract}
We study molecular emission in a massive condensation at the border of the \hii{} region RCW~120, paying particular attention to the Core~1 and Core~2 objects, the most massive fragments  of the condensation found previously by ALMA. The latter fragment was previously suggested to host a high-mass analogue of Class 0 young stellar object. We present spectra of molecular emission in the 1~mm range made with the APEX telescope. We detect CH$_3$OH and \cs{} lines in Core~1 and Core~2. The CH$_3$CN series and the SO$_2$ lines are only found in Core~2. We estimate gas physical parameters using methanol lines and obtain gas temperature less than 100~K in both regions. Molecular hydrogen number density in Core~2 is in the range of $10^5-10^7$~cm$^{-3}$ and is more uncertain in Core~1. However, the detection of the CH$_3$CN lines corresponding to highly excited transitions ($E_{\rm u}> 400$~K) in Core~2 indicates that the region contains hot gas, while the abundances of  CH$_3$OH, CS,  SO$_2$ and CH$_3$CN are quite low for a hot core stage. We propose that Core~2 is in the warm-up phase prior to the establishing of the hot gas chemistry. We suggest that Core~2 is in the beginning of the hot core stage. There are no detected CH$_3$CN lines in Core~1, therefore, it might be on an even less evolved evolutionary stage.
\end{abstract}


\section{Introduction}

The initial stage of high-mass star formation is a subject of debates. While starless dense clumps of molecular clouds, precursors of protostellar cores for low-mass stars, were observed by, for example,~\citet{Tafalla_1998}, similar high-mass objects are rare and more difficult to find and confirm~\citep[see, e. g. modelling by][]{Pavlyuchenkov_2011}. Several candidates were proposed by \citet{2005ApJ...634L..57S, 2010ApJ...715.1132O, 2012MNRAS.423.2342F, 2019ApJ...886...36S, 2020arXiv201207738Z}. One of the most massive and well-known dense gas condensations at the border of the \hii{} region RCW~120 has been considered as a high-mass analogue of a Class~0 object for about ten years since the studies by \citet{Deharveng_09, Zavagno_2010}. In particular, \citet{Deharveng_09} found a compact 870~\mkm{} core of 250~M$_\odot$ adjacent to the ionization front of RCW~120, which is embedded into a 800-M$_\odot$ gas condensation at the south-western border of RCW~120 (Clump~1 hereafter). \citet{Zavagno_2010}, using {\it Herschel} infrared data, confirmed Class~0 properties of the object and found that the emission towards the dense core is dominated by its 10$^3$~M$_\odot$ envelope. The formation of the gas condensation with the massive dense core at the border of the \hii{} region was proposed as an outcome of a triggering collect-and-collapse (C\&C) process related to expansion of RCW~120 in the studies by \citet{Zavagno_2007, Deharveng_09, Minier_2013}. \citet{Tremblin_13} suggested that compression by an ionization front played a major role, along with gravity, for the dense core formation in Clump~1. Recently, \citet{Figueira_2020} re-examined the triggered star-formation model with new observations of CO lines made by the APEX telescope, and confirmed that it might be at work at the edges of RCW~120. \citet{Zavagno_2020} found a filamentary structure of a photodissociation region around RCW~120 using ArT\'{e}MiS on APEX. They concluded that the compression of pre-existing molecular filaments in Clump~1 may influence star-forming regions there.

\citet{Figueira_2017} re-examined the {\it Herschel} data and identified 35 compact young stellar objects (YSOs) with reliable SEDs associated with RCW~120. They found that four out of five massive YSOs near RCW~120 are located in Clump~1. Analysing the ALMA data, \citet{Figueira_2018} found that the most massive YSO \citep[Core~2 hereafter, an object with ID~2 in Table~5 in][]{Figueira_2017} probably associated with an ultra-compact \hii{} region (uc\hii). They reported the detection of CH$_3$CN and SO$_2$ emission at 3~mm but did not analyse this emission in detail and did not determine the physical properties of the gas in the cores. 

In this paper, we analyse data on molecular emission in Clump~1 at 1~mm in order to determine physical parameters of the gas in Core~2, and also in Core~1, which is the second most massive YSO after Core~2 in Clump~1. We focus mainly on methanol and CH$_3$CN emission, as these molecules are reliable tracers of physical conditions in molecular gas. Methanol is an abundant, well-known interstellar molecule, especially in regions of star formation \citep[see, e. g.][]{Kalenskii_2002, Leurini_2007}. Being a slightly asymmetric rotor, methanol has a complex spectrum, sensitive to the physical conditions in dense interstellar gas \citep{Kalenskii_1994, Leurini_2004, Salii_2006}. The CH$_3$CN transitions are used to obtain independent measurements of gas kinetic temperature, as has been shown, for example by~\citet{Kalenskii_2000,Gratier_2013,Beltran_2018}. We describe our observations in Sec.~\ref{Sec:Obs}; we present our results (physical properties of the gas and molecular abundances) and analysis in Sec.~\ref{Sec:res}. Our theoretical modelling of the chemical evolution of Core~2 is presented in Sec.~\ref{sec:model}; those results are discussed in Sec.~\ref{sec:disc}. The summary of our study is presented in Sec.~\ref{sec:conc}.

\section{Observations and data analysis}\label{Sec:Obs}

In July 2009, using APEX\footnote{This publication is based on the data acquired with the Atacama Pathfinder Experiment (APEX). APEX is a collaboration between the Max-Planck-Institut fur Radioastronomie, the European Southern Observatory, and the Onsala Space Observatory.}, we observed several selected positions in Clump~1 around the maximum of 870~$\mu$m emission found by \citet{Deharveng_09}: $\alpha_{2000}=17^{\rm h} 12^{\rm m} 08^{\rm s}$ and $\delta_{2000}=-38^\circ$30\arcmin\,45\arcsec\, (project number O-083.F-9311A-2009). Several selected positions and methanol spectra at 241791.431~MHz are shown in Fig.~\ref{fig:meth_241791_around}. Relative coordinates of the all observed positions are given in Table~\ref{tab:observedpositions}. The observations were carried out in service mode in good weather conditions, with  a  Precipitable Water Vapor (PWV) amount between 0.4 and 0.7~mm. Two spectral windows\footnote{The observations in one more window, centered at $330$~GHz, were presented by \citet{Kirsanova_2019} and will not be discussed here.}, at $\sim 220$~and $\sim 241$~GHz, were observed with the SHeFI receiver APEX-1 \citep[][]{Belitsky_2006, Vassilev_2008}. The spectrometer consisted of two FFTS units configured to provide a spectral resolution of 122~kHz (0.17~\kms) and to cover frequency bands of 1400~MHz for the 220~GHz spectral window and 1200~MHz for the 241~GHz window (0.16~\kms). The main instrumental parameters are summarised in Table~\ref{tab:obs}. The data reduction was performed using the GILDAS software\footnote{\url{http://www.iram.fr/IRAMFR/GILDAS}}. The identification of the detected lines was performed using the CDMS database \citep{Endres_2016}.

\begin{table}
\centering
\caption{Basic observational parameters.}
\begin{tabular}{cccc}
\hline
Reciever & Freq. range   & Beam      & rms\\
         & (GHz)         & (\arcsec) & (K)\\
\hline
APEX-1   & 220.0--221.4  & 29        & 0.08-0.14\\
APEX-1   & 240.7--241.9  & 26        & 0.09-0.15\\
\hline
\end{tabular}
\label{tab:obs}
\end{table}

The observed methanol line series allow us to estimate excitation conditions and line opacities. We used large velocity gradient (LVG) analysis to determine the physical parameters of gas where the methanol emission appears, namely: gas kinetic temperature (\tgas, K), hydrogen number density ($n_{\rm H_2}$,~cm$^{-3}$), methanol specific column density ($N_{\rm CH_3OH}/\Delta V$,~cm$^{-3}$~s) and methanol relative abundance ($N_{\rm CH_3OH}/N_{\rm H_2}$). Since the beam filling factor for the methanol emission ($f$) is not known, we include this additional parameter in the methanol line intensity analysis. The details of the LVG calculation and a database of population numbers for the methanol energy levels, calculated for several values of line width (1, 3 and 5~\kms) can be found in, for example, the work by \citet{Kirsanova_2017,Salii_2018}. In the present calculations, we use a line width of 5~\kms{} for the detected methanol lines since this is the closest value to the observed widths given in Table~\ref{tab:detected}. Dust emission and absorption within the emission region are taken into account, as described by \citet{Sutton_2004}, where dust and gas temperatures were set to be the same. Using this approach, we can obtain a fit with the best coincidence between the modelled and observed line intensities (with $\chi^2$ minimum). The upper limits of the undetected methanol lines were included in the consideration: it was controlled that their model intensities not exceed the rms level from Table~\ref{tab:obs}.

To extend the analysis of the methanol emission and estimate the parameters in the regions with the weaker lines, we applied a Bayesian analysis \citep[e.g.][]{Ward2003} to obtain the confidence intervals. We calculated methanol model intensities ($T_{{\rm m}\,i}(p)$) for a regular network of parameters: $$ p = \left(T_{\rm gas} , N_{\rm CH_3OH}/\Delta V, n_{\rm H_2}, N_{\rm CH_3OH}/N_{\rm H_2}, f\right)$$ using the database of the population numbers for methanol quantum energy levels.

The probability to observe $N$ methanol lines with intensities $T_{{\rm o}\,i}$ and uncertainties $\sigma_i$ using a set of parameters $p$ can be calculated as:
\begin{equation}
P(T_{\rm o}|p) = \prod_i^N{\frac{1}{\sqrt{2\pi}\sigma_i}\,e^{-\frac{1}{2}\left(\frac{T_{{\rm o}\,i} - T_{{\rm m}\,i}(p)}{\sigma_i}\right)^2}}.
\end{equation}
Integrating over each parameter, we obtain the Bayesian probability function and estimate the parameters and confidence intervals.

The analysis of the CH$_3$CN line emission was done with the population diagram method~\citep[see, e.g.][]{Goldsmith_1999, Kalenskii_2000}. The column densities of CS and SO$_2$ molecules were determined under LTE assumption using a standard approach described by \citet{Mangum_2015} and using the gas kinetic temperature derived from the methanol emission analysis as the excitation temperature in the LTE approach. We recognise all the restrictions of the LTE approach and consider the LTE analysis only as a way to obtain estimates of molecular column densities since we have only one line for each molecule. To convert column density of \cs{} to CS, we used an elemental abundance ratio [S]/[$^{34}$S]~=~22.5~\citep{Wilson_1999}.

In order to calculate the molecular abundances relative to molecular hydrogen from their column densities, we use the column density of H$_2$ molecules $N_{\rm H_2} = 3.2_{1.6}^{4.8}\times 10^{23}$\,cm$^{-2}$ obtained in the ViaLactea project \citep{2015MNRAS.454.4282M, 2017MNRAS.471.2730M}, which is also in agreement with the column density values obtained by \citet{Anderson_2012, Tremblin_13} and \citet{Figueira_2017}.

\section{Results}\label{Sec:res}

\begin{figure*}
\includegraphics[width=15cm]{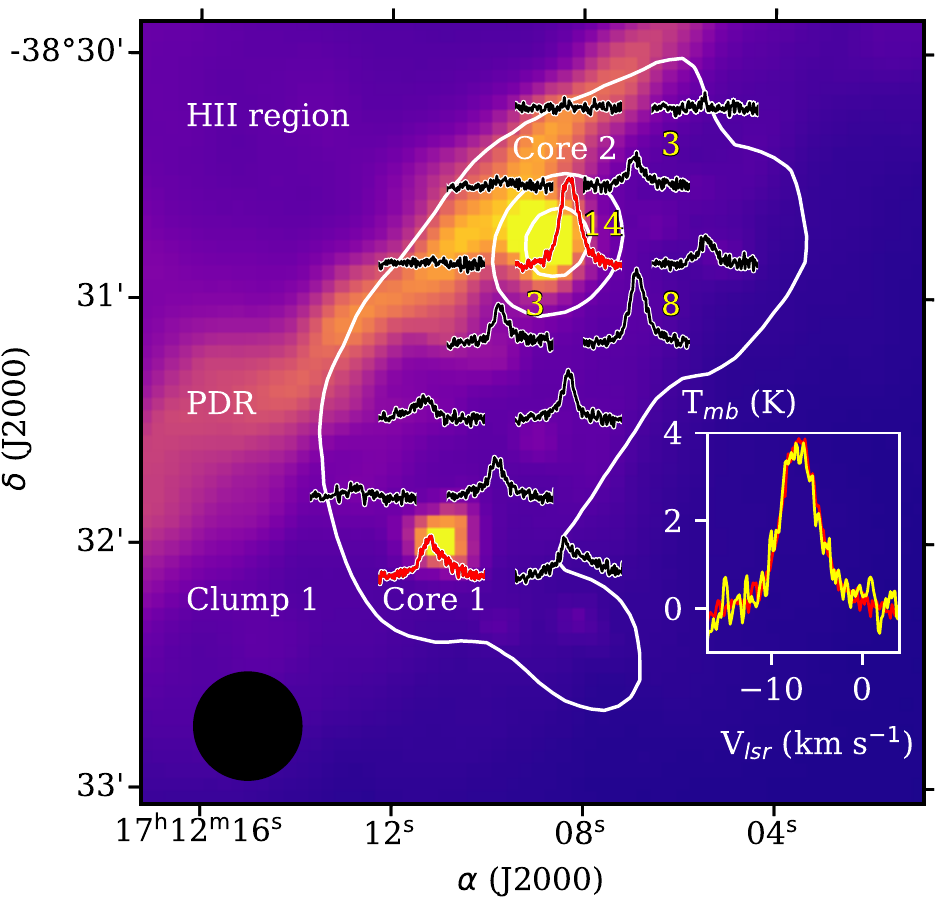}
\caption{{\it Herschel} image of Clump~1 at 70~$\mu$m. White contours represent the 870~\micron{} ATLASGAL contours for 1, 5 and 10 Jy~beam$^{-1}$. Methanol spectra at 241791.431~MHz in the observed positions are shown by black lines. The spectra towards Core~1 (southern part of Clump~1) and Core~2 (northern part of Clump~1) are shown by red colour. The methanol line (red), overlaid upon the \cs(5--4) line (yellow, multiplied by a factor of 20), towards Core~2 are also shown in the separate bottom frame. The overall brightness temperature and velocity scales are shown on this frame. The numbers above four methanol spectra represent $N_{\rm CS}$ divided by $10^{13}$. The APEX beam of 26\arcsec{} at 241 GHz is shown by the black circle.}
\label{fig:meth_241791_around}
\end{figure*}

The full list of the detected lines in Cores~1 and 2 and the parameters of Gaussian fits are shown in Table~\ref{tab:detected}. The spectra of the methanol and CS line emission are shown in Fig.~\ref{fig:meth_241791_around},~\ref{fig:meth_sp_S1S2} and~\ref{fig:meth_examples}, while the spectra of CH$_3$CN are shown in Fig.~\ref{fig:ch3cndiagr}. We note that the lines with an upper level energy $E_{\rm u} > 50$~K are detected only towards Core~2. Three lines with $T_{\rm mb} \approx 0.5-1$~K around 220050, 241895 and 240903~MHz remain unidentified in Fig~\ref{fig:meth_sp_S1S2}. The brightest among of them at 220050~MHz is detected in both Core~1 and 2 with $T_{\rm mb} \approx 1$~K.

\begin{table*}
\centering
\caption{Detected molecular lines towards Core~1 and Core~2 in RCW~120. We used single-peaked Gaussian profiles to fit the lines. $^*$Velocities and widths of the $K \le 3$ CH$_3$CN lines are assumed to be the same in the fitting. $^{**}$Velocities and widths of the $K \ge 4$ CH$_3$CN lines are also assumed to be the same. The methanol lines at 241832.716 and 241833.104~MHz are not resolved in our spectrum, therefore, their parameters are set to be same in our analysis. The same is for the lines at 241842.324 and 241843.646~MHz. Spectroscopic entries are taken from the CDMS database: C$^{34}$S, ver.~2, Jan. 2004;   SO$_2$, ver.~2, July 2005; CH$_3$OH, ver.~3, May 2016;  CH$_3$CN, ver.~2, Nov. 2016.}
\begin{tabular}{lccccccc}
\hline
Sym                    & Transition                                &
Frequency                  & $E_{\rm u}$& $\int T_{\rm mb} dV$ 
&$V_{\rm LSR}$ &$\Delta V$   & T$_{\rm mb}$ \\
                        &                                           &
[MHz]                      & [K]            & [K\,\kms]            &
[\kms]       &[\kms]       & [K]          \\
\hline
\multicolumn{8}{l}{\it Core~1}\\
                        \multicolumn{8}{c}{\bf \meth}\\
E &$5_0-4_0 $ &241700.159 &47.9& $ 3.01\pm 0.20$ &$-6.66\pm 0.17$ &$5.27\pm 0.44$ &$0.54\pm 0.06$\\

E & $5_{-1}-4_{-1} $& 241767.234 & 40.4 & $10.74\pm 0.23$&$-7.23\pm 0.06$ &$6.00\pm 0.17$ &$1.68\pm 0.06$\\
A$^+$& $5_{0}-4_{0} $ & 241791.352  & 34.8 & $12.70\pm 0.25$&$-7.24\pm 0.06$ &$6.36\pm 0.16$ &$1.88\pm 0.06$\\
\multicolumn{8}{c}{ }\\
\multicolumn{8}{l}{\it Core~2}\\
                        \multicolumn{8}{c}{\bf \cs}\\
--   &   $5-4$&241016.089&27.8&$ 3.99\pm 0.12$&$-6.38\pm 0.06$ &$4.69\pm 0.16$ &$0.81\pm 0.04$\\
\multicolumn{8}{c}{ }\\
\multicolumn{8}{c}{\bf \meth}\\
E&$8_0-7_1$&220078.561 & 96.6& $ 1.14\pm 0.15$ &$-6.65\pm 0.33$ &$5.11\pm 0.79$ &$0.21\pm 0.04$\\
E&$5_0-4_0 $&241700.159 &47.9& $ 6.98\pm 0.21$&$-6.33\pm 0.06$ &$3.90\pm 0.14$ &$1.68\pm 0.08$\\
E&$5_{-1}-4_{-1} $&241767.234&40.4& $14.85\pm 0.19$&$-6.41\pm 0.03$ &$4.38\pm 0.07$ &$3.19\pm 0.07$\\
A$^+$&$5_{0}-4_{0} $&241791.352&34.8& $16.00\pm 0.19$&$-6.25\pm 0.03$ &$4.38\pm 0.06$ &$3.43\pm 0.07$\\
A$^+$ &$5_3-4_3$&241832.718&84.6& $ 0.31\pm 0.14$&$-6.96\pm 0.57$ &$2.21\pm 0.76$ &$0.13\pm 0.09$\\
A$^-$&$5_3-4_3$&241833.106&84.6& -"-&-"- &-"- &-"-\\
A$^-$&$5_2-4_2$&241842.284&72.5& $ 0.56\pm 0.17$&$-6.74\pm 0.56$ &$3.45\pm 0.92$ &$0.15\pm 0.09$\\
E&$5_3-4_3$&241843.604&74.6&  -"-&-"-&-"-&-"-\\
E&$5_1-4_1$&241879.025&55.9& $ 1.66\pm 0.17$&$-6.24\pm 0.14$ &$2.87\pm 0.32$ &$0.55\pm 0.09$\\
A$^+$&$5_2-4_2$&241887.674&72.5& $ 0.23\pm 0.19$&$-6.29\pm 1.03$ &$2.23\pm 2.22$ &$0.10\pm 0.09$\\
\multicolumn{8}{c}{ }\\
                        \multicolumn{8}{c}{\bf CH$_3$CN}\\
                 A      & $12_0-11_0 $                              &
220747.261                 & 68.9                 &$1.29\pm 0.11$  &
$-6.60\pm 0.17$         &$4.81\pm 0.22$           &$0.25\pm 0.02$\\
                 E      & $12_1-11_1 $                              &
220743.011                 & 76.0                 &$1.28\pm 0.11$  &
$-6.60\pm 0.17$         &$4.81\pm 0.22$           &$0.25\pm 0.02$\\
                 E      & $12_2-11_2 $                              &
220730.261                 & 97.4                 &$0.89\pm 0.12$  &
$-6.60\pm 0.17$         &$4.81\pm 0.22$           &$0.18\pm 0.02$\\
                 A      & $12_3-11_3 $                              &
220709.017                 & 133.6                &$0.87\pm 0.11$
&$-6.60\pm 0.17$        &$4.81\pm 0.22 $          &$0.17\pm $ 0.02\\
                 E      & $12_4-11_4 $                              &
220679.287                 & 175.1                &$0.68\pm 0.14$
&$-5.22\pm 0.59 $       &$9.11\pm 0.84 $          &$0.07\pm $ 0.02 \\
                 E      & $12_5-11_5 $                              &
220641.084                 & 247.6                &$0.40\pm 0.14$  &
$-5.22\pm 0.59 $        &$9.11\pm 0.84 $          &$0.04\pm 0.02$ \\
                 A      & $12_6-11_6 $                              &
220594.424                 & 326.2                &$0.56\pm 0.15$  &
$-5.22\pm 0.59 $        &$9.11\pm 0.84 $          &$0.06\pm 0.02$\\
                 E      & $12_7-11_7 $                              &
220539.324                 & 418.9                &$0.47\pm 0.17$  &
$-5.22\pm 0.59 $        &$9.11\pm 0.84 $          &$0.05\pm 0.02$ \\
                        \multicolumn{8}{c}{ }\\
                        \multicolumn{8}{c}{\bf SO$_2$}\\
  --  &$5_{2,4}-4_{1,3} $
&241615.797                  &23.6                   &$0.57\pm 0.13$
&$-6.10\pm 0.50 $&$4.27\pm 1.02 $&$0.13\pm 0.03$\\

\hline
\end{tabular}
\label{tab:detected}
\end{table*}

\subsection{Methanol emission towards Cores~1 and 2}
 
\begin{figure*}
\includegraphics[width=1.99\columnwidth]{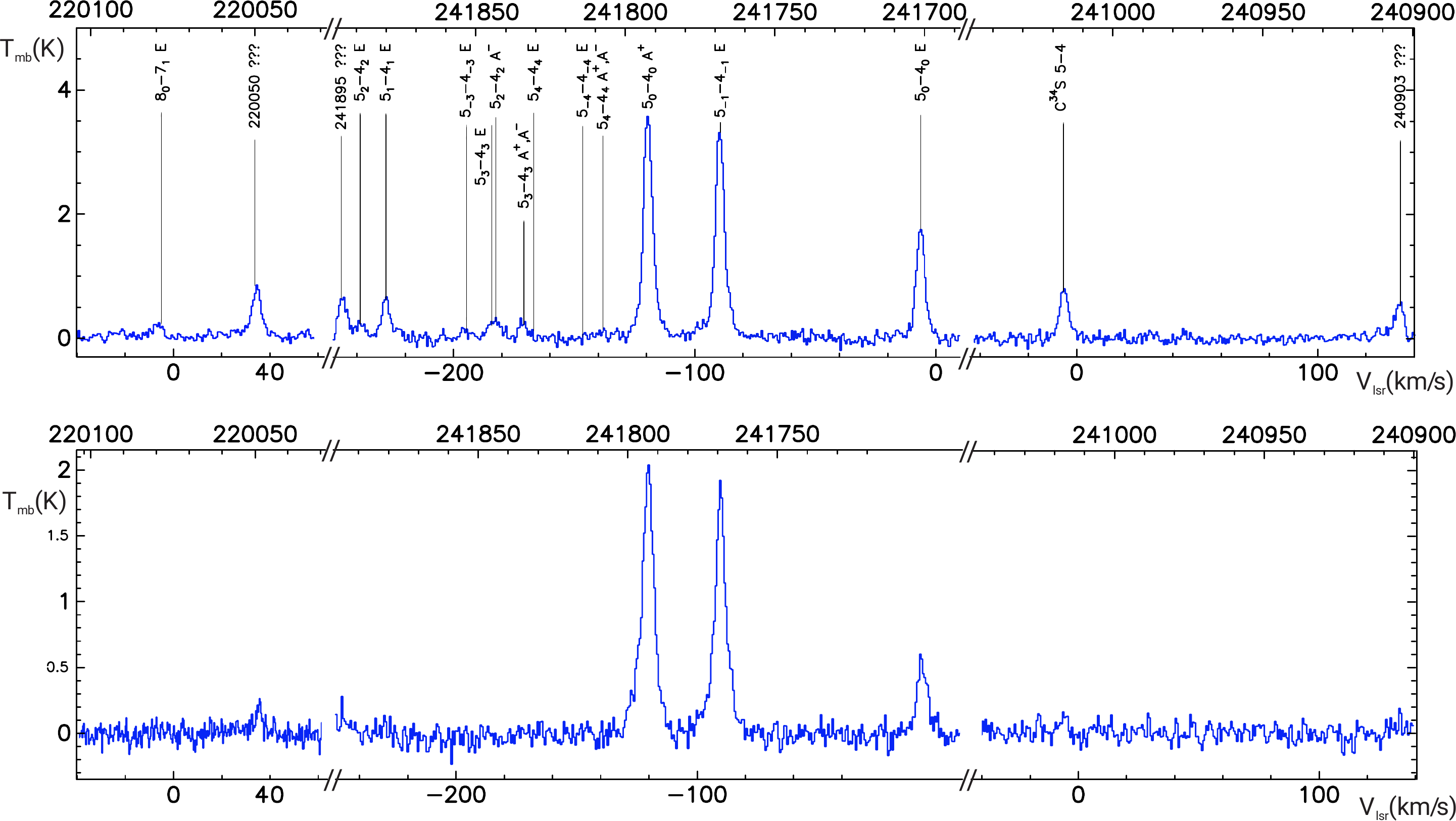}
\caption{Detected lines at 220 and 241~GHz towards Core~2 (top) and 1 (bottom)}
\label{fig:meth_sp_S1S2}
\end{figure*}

The methanol lines are bright both in Core~1 and in Core~2, but the brightest lines are detected in Core~2 (see Fig.~\ref{fig:meth_sp_S1S2}). We note that only lines with $E_{\rm u} < 70$~cm$^{-1}$ ($\approx 100$~K) were confidently (at level above $3 \sigma$) detected in the unsmoothed spectra. Only three lines with $E_{\rm u} \leq 50$~K were detected at a level above $3 \sigma$ in Core~1. All the detected methanol lines have single-peaked profiles.

\begin{figure}
\includegraphics[width=0.99\columnwidth]{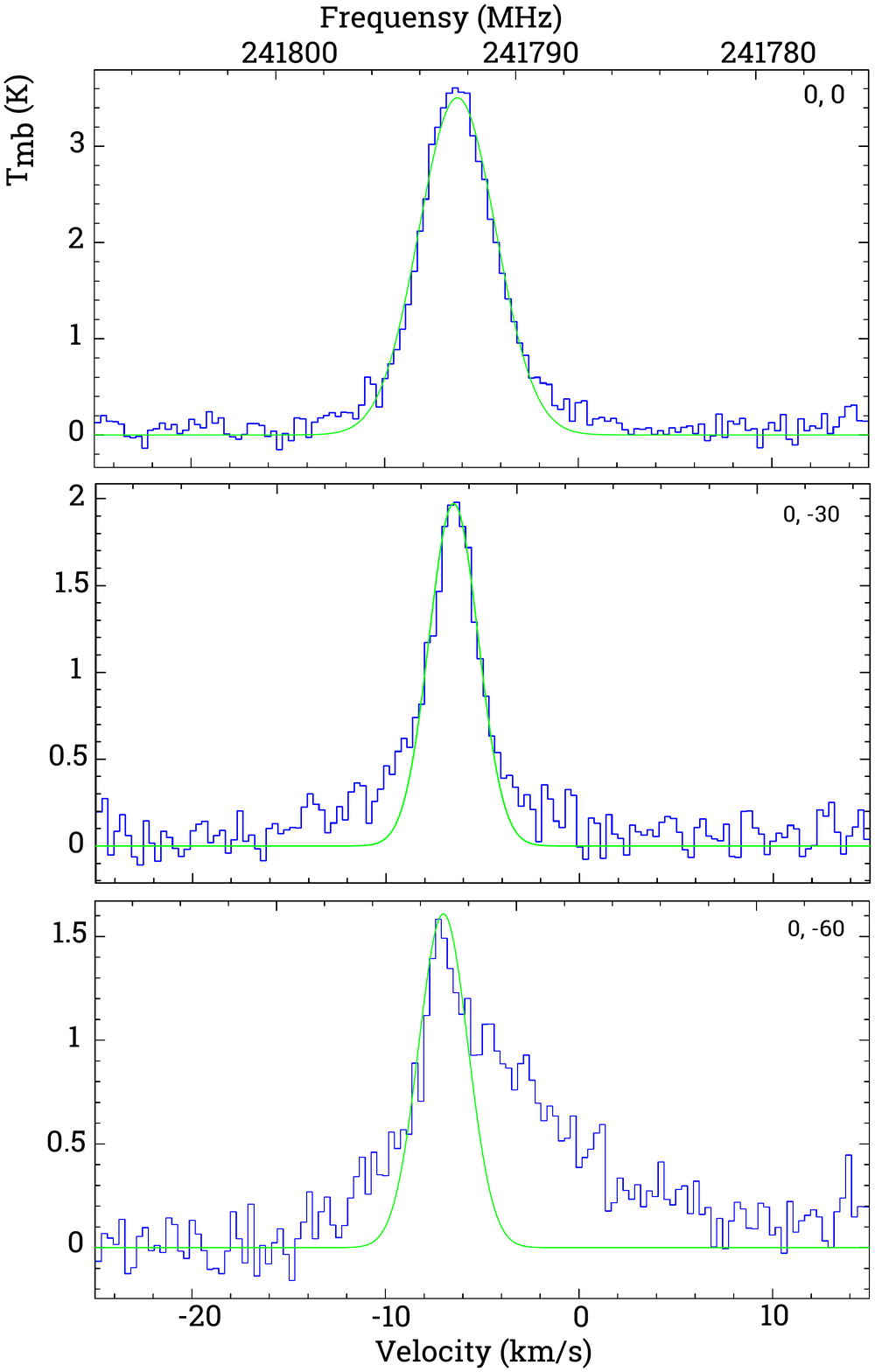}
\caption{Examples of methanol lines at 241791.431~MHz towards Core~2 (top) and two positions 30 and 60\arcsec{} to the south from the Core~2. Fitting by single-peak Gauss profile is shown by green. Parameters of the fit towards Core~2 are given in Table~\ref{tab:detected}.}
\label{fig:meth_examples}
\end{figure}

The brightest methanol lines at 241700, 241767 and 241791~MHz in Core~2 have a Gaussian shape with red (about 4~\kms) and blue (about 2~\kms) wings at 15\% level from the intensity maximum. The line wings in Core~2 are better seen in Fig.~\ref{fig:meth_examples}. The wings also can be fitted by Gaussian profiles with centres close to the main line centres, but with widths about 3--5 times larger than the main line. The methanol emission fades quickly to the north and west and stretches to the south and south-east following the distribution of \co{} emission presented by \citet{Kirsanova_2019} and \citet{Figueira_2020}. There is a negative velocity gradient of the methanol lines from Core~2 to Core~1, similar to the \co{} and \cvo{} lines \citep{Kirsanova_2019, Figueira_2020}, and almost no velocity gradient perpendicular to the dense shell. The methanol spectra at Core~1 also have pronounced red and blue wings. The most prominent wings are found to the west of Core~1 (see Fig.~\ref{fig:meth_examples}), where the integrated intensity of the red wing is comparable with the intensity under the Gaussian profile centered at the peak intensity.

\begin{figure}
\includegraphics[width=0.99\columnwidth]{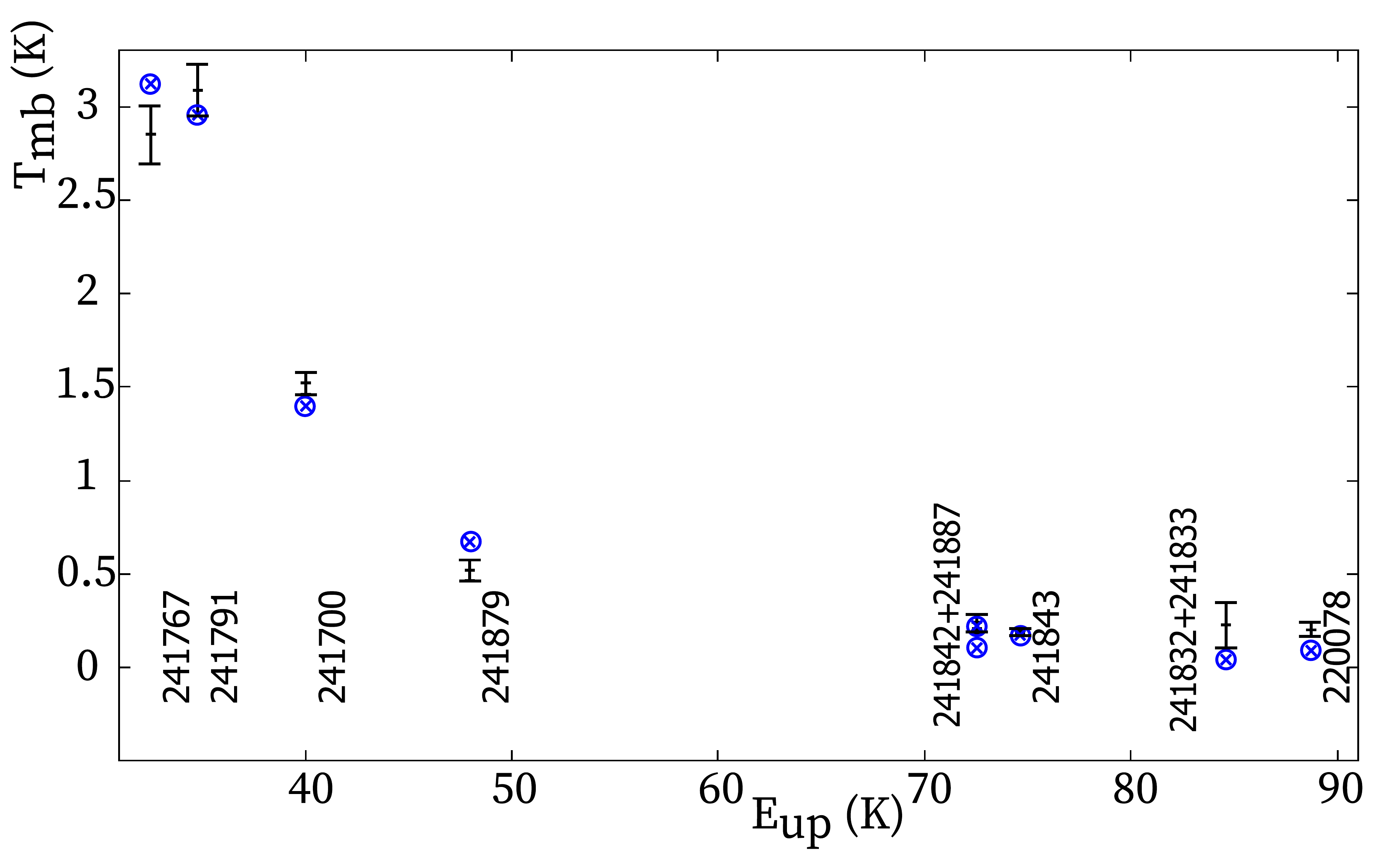}
\caption{Results of the LVG modelling. Blue symbols show results of the modelling, black symbols with errorbars show observed brightness temperatures of the methanol lines. Frequencies of the used lines are given in MHz.}
\label{fig:LVGres}
\end{figure}

Our LVG analysis of methanol emission, described above, allows us to conclude that all the detected methanol lines have modest optical depth. In particular, the brightest methanol lines at 241700 and 241791~MHz have optical depths around 0.8. For Core~2, we obtain the best agreement between the model and observed line intensities for the parameters: $T_{\rm gas} = 30_{20}^{40}$~K, $N_{\rm CH_3OH}/\Delta V = 5.0_{3.9}^{13.0}\times 10^9$~cm$^{-3}$\,s, $n_{\rm H_2} = 1.0_{0.3}^{3.1} \times 10^6$~cm$^{-3}$ and a relative abundance of methanol $x_{\rm CH_3OH} = 3.2_1^{316} \times 10^{-9}$ and $f = 50_{20}^{60}$\%. A comparison between the observed and best-fit model line brightness temperatures is shown in Fig.~\ref{fig:LVGres}.
The number of the detected methanol lines towards Core~1 is insufficient to estimate the parameters using the $\chi^2$ fit with the LVG method. Therefore, we only estimated the Bayesian confidence intervals for the parameters. These intervals are $T_{\rm gas} < 60$\,K, $N_{\rm CH_3OH}/\Delta V = (7.9-20)\times 10^9$~cm$^{-3}$\,s, $n_{\rm H_2} = 3.2\times 10^3 - 6.3\times 10^6$~cm$^{-3}$,  $N_{\rm CH_3OH}/N_{\rm H_2} > 10^{-8}$, $f>10$\% for a $1\sigma$ significance level.

\subsection{CH$_3$CN emission towards Core~2}\label{sec:ch3cn}

\begin{figure}
\includegraphics[width=\columnwidth]{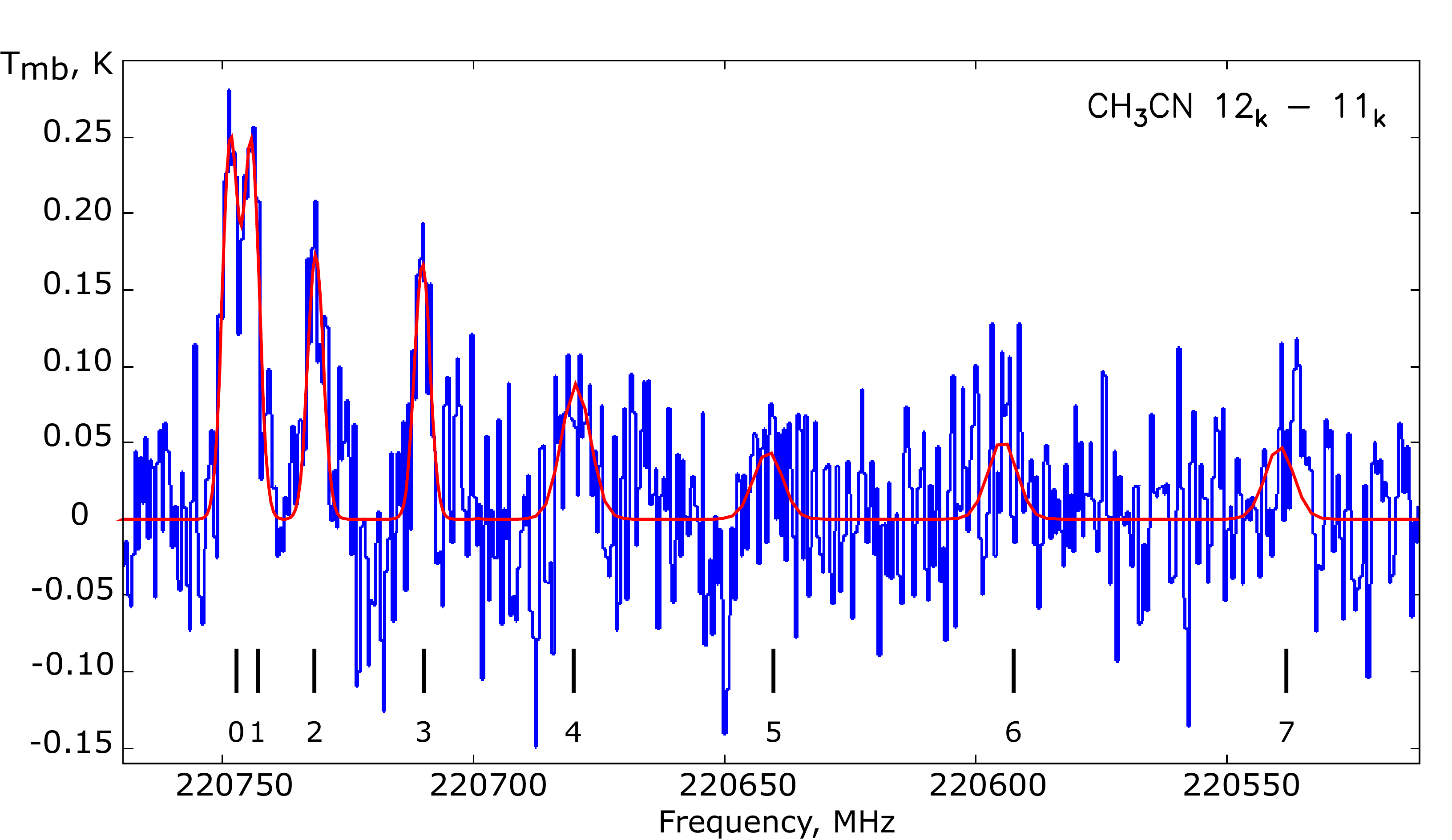}
\includegraphics[width=\columnwidth]{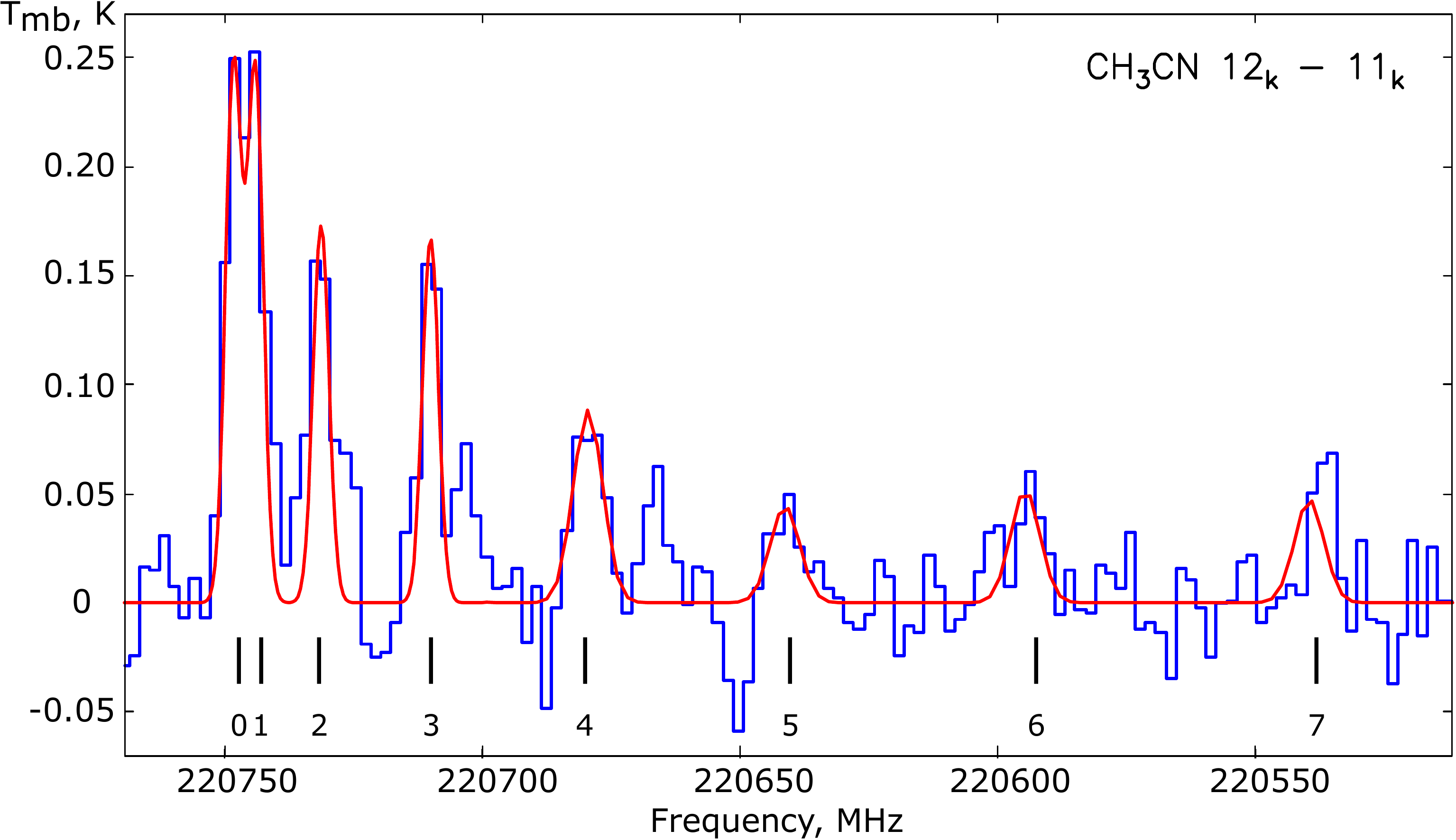}
\includegraphics[width=\columnwidth]{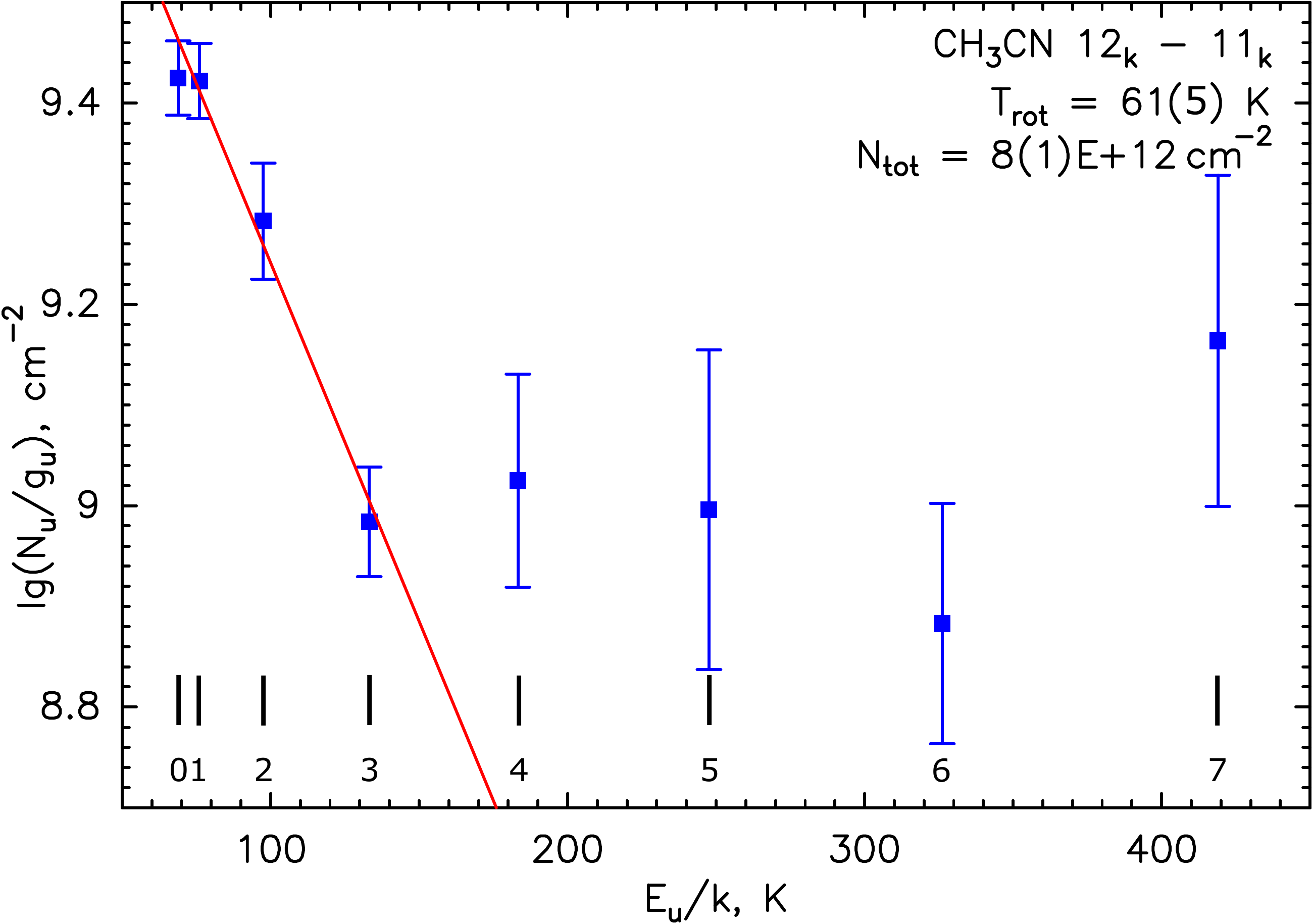}
\caption{Top: original CH$_3$CN(12$_K$--11$_K$) spectrum towards Core~2; middle: twice Hanning smoothed original
spectrum. The Gaussian fits of the $K=0-7$ lines presented in Table~\ref{tab:detected} are overlayed on both spectra; bottom: rotation diagram with a linear fit over four $K=0, 1, 2, 3$ lines.}
\label{fig:ch3cndiagr}
\end{figure}

We find eight $12_K-11_K$ CH$_3$CN lines towards Core~2. Four lines with excitation energies below 140~K, corresponding  to $K=0, 1, 2, 3$, are visible with the original spectral resolution (Fig.~\ref{fig:ch3cndiagr}, upper panel). In comparison, four lines with excitation energies above 140~K ($K=4, 5, 6, 7$) are clearly seen only on the spectrum, where the resolution is reduced to 1.3~\kms{} by double Hanning smoothing (Fig.~\ref{fig:ch3cndiagr}, middle panel). As the widths of the low-excitation and high-excitation lines are different, we fitted them separately. First, the low-excitation lines were fitted, setting their velocities and widths to be the same, and then the same procedure was applied to the high-excitation lines. For the low-excitation lines, we obtained $V_{\rm LSR}=-6.6$~\kms{} and $\Delta V=4.8$~\kms{}, which is close to the same parameters of the methanol lines. The high-excitation lines are broader, $\Delta V=9.1$~\kms{}, and the central line velocity is slightly different, $V_{\rm LSR}=-5.2$~\kms{}. 

The four detected low-excitation lines allow us to estimate the CH$_3$CN excitation temperature using the rotational diagram method. The points corresponding to their upper-level populations lie on a straight line (Fig.~\ref{fig:ch3cndiagr}, lower panel).  We find $T_{\rm ex}=61$~K and column density $N_{\rm CH_3CN} = 8.1\times10^{12}$~cm$^{-2}$ applying an optically thin analysis. We consider these lines as optically thin, because their intensity becomes lower for higher $E_{\rm up}$ values. This is in contrast, for example, with excitation of the lower~K transitions in a recent study by \citet{2020ApJ...904..181L}. Since the LSR velocities and widths of low-excitation CH$_3$CN lines coincide well with those of the CH$_3$OH lines, most likely both of them arise in the same region. Using $N_{\rm H_2} = 3.2\times 10^{23}$\,cm$^{-2}$, as above, we obtain 
an abundance of CH$_3$CN $x_{\rm CH_3CN} = 2.5\times 10^{-11}$ in the low-excitation molecular gas towards the centre of Core~2. This value is typical for dense cores in massive star-forming regions~\citep{Kalenskii_2000}. The relative abundance is $[{\rm CH_3CN/CH_3OH}]=N_{\rm CH_3CN}/N_{\rm CH_3OH}\approx 10^{-2}$. 

The points corresponding to the spectral lines with $K \geq 4$ do not lie on the straight line approximating the low-excitation level populations. The fact that the points are located above the approximating line demonstrates excess emission in the high-excitation lines, which, in turn, implies the presence of hot gas in Core~2. These points do not lie on any straight line, which can be naturally explained, assuming that the high-excitation lines are not optically thin. Note that the point corresponding to the doubly degenerate $K=6$ line jumps down with respect to its neighbors, and such behavior of the degenerate lines is an indicator of high optical depth~\citep{Kalenskii_2000, Remijan_2004, Beltran_2018}. The observed high-excitation lines are very weak, and if they are optically thick, the hot region must be very compact. We estimated $T_{\rm ex}$ assuming an optically thin approximation for the lines with $K \geq 4$ in order to consider all possible scenarios, but did not obtain a statistically significant result for the $T_{\rm ex}$ value, due to the weakness of the lines. Therefore, additional observations are needed in order to confidently determine the gas temperature in Core~2. We should nevertheless point out that the high-excitation CH$_3$CN lines in star-forming regions trace hot cores~\citep{1996A&A...315..565O, Kalenskii_2000, Remijan_2004, Beltran_2018}, therefore we suggest that Core~2 is at the very beginning of the hot core stage. 

Neither low-excitation nor high-excitation CH$_3$CN lines are detected in Core~1.

\subsection{Emission of \cs{} and SO$_2$ molecules in Core~2}

The spectrum of the \cs(5--4) line towards Core~2 is shown in Fig.~\ref{fig:meth_241791_around}. The line profile is similar to that of the methanol line at 241791.431~MHz (we multiply the \cs(5--4) profile by a factor of 20 to show this in Fig.~\ref{fig:meth_241791_around}). Therefore, the methanol and \cs(5--4) emission mainly originates in the same region. We found \cs(5--4) lines not only directly towards Core~2, but also in the surrounding area with a radius $\approx 15-30$\arcsec{} (0.1--0.2~pc at a distance of 1.34\,kpc~\citet{Russeil_03}). Assuming LTE conditions and using $T_{\rm ex}=30$~K, $T_{\rm bg}=2.7$\,K, $\mu = 1.96\times 10^{-18}$~esu, $B_0 = 24495.576$~MHz, $E_{\rm u} = 27.8$~K, $J_{\rm u} =5$ and filling factor $f=0.4$ (from the methanol emission analysis), we calculate the column density of CS ($N_{\rm CS}$). The parameters of the \cs(5--4) lines are shown in Table~\ref{tab:detected}. The values of $N_{\rm CS}$, obtained as $N_{\rm CS} = 22.5 N_{\rm C^{34}S}$, in Core~2 and around are shown in Fig.~\ref{fig:meth_241791_around}. The relative abundance of CS, $x_{\rm CS}$, towards Core~2 is $4.5\times 10^{-10}$. The value of $N_{\rm CS}$ is about 2 times lower at an offset of 15\arcsec{} offset in the direction perpendicular to the ionization front (14 vs $8\times10^{13}$~cm$^{-2}$) and 4.5 times lower in two 15\arcsec{} offsets along the front of the photodissociation region, visible as a bright bar in Fig.~\ref{fig:meth_241791_around}. Bearing in mind the high critical density of the \cs(5--4) transition, which is higher than 10$^6$~cm$^{-3}$ for gas temperatures from 10 to 100~K, we conclude that the dense gas distribution in Core~2 is asymmetric and elongated in the south-west direction from the centre of Core~2.

Our frequency setup covers the SO$_2$($5_{2,3}-4_{1,3}$) line at 241615.779~MHz, and we find this line in the spectrum towards Core~2 after smoothing by factor of 4. The parameters of the detected SO$_2$(5$_{2,4}$--4$_{1,3}$) line towards Core~2 are shown in Table~\ref{tab:detected}.  Assuming LTE conditions with $T_{\rm ex}=30$~K, as we used for the analysis of \cs{} emission, $T_{\rm bg}=2.7$\,K, $\mu = 1.63\times 10^{-18}$~esu, rotational constants $A_0 = 60788.550$~MHz, $B_0 = 10318.074$~MHz, $C_0 = 8799.703$~MHz, $E_{\rm u} = 23.6$~K, filling factor $f=0.4$ (from the methanol emission analysis) and using Eq.~57 from  \citet{Mangum_2015} for the asymmetric molecule partition function, we obtain SO$_2$ column density $N_{\rm SO_2} = 2.32\times10^{12}$~cm$^{-2}$ and relative abundance $x_{\rm SO_2}$ in Core~2 of $6.6\times 10^{-12}$.

\section{Astrochemical modelling}\label{sec:model}

In order to check whether our abundance estimates are viable for any particular stage of the star-formation process, we performed a simple astrochemical analysis relying upon the physical parameters found above for Core~2. The astrochemical model PRESTA \citep{presta} is used for that purpose. This model solves the chemical kinetic equations accounting for both gas-phase and solid-phase processes, which are taken from the ALCHEMIC database \citep{alchemic} with additions described by \cite{wiebe2019}. Abundances and physical constraints are summarised in Table~\ref{tab:chemmodel}. We note that chemical modelling procedure normally uses number density of hydrogen atoms $n_{\rm H} = 2 n_{\rm H_2}$.

\begin{table}
\centering
\caption{Parameters of the astrochemical model for Core~2.}
\begin{tabular}{cc}
\hline
parameter & value \\
\hline
$n_{\rm H}$ & $10^4-10^6$~cm$^{-3}$ \\
$x_{\rm CS}$ & $4.5\times10^{-10}$ \\
$x_{\rm SO_2}$ & $6.6\times10^{-12}$ \\
$x_{\rm CH_3OH}$ & $3.2\times10^{-9}$ \\
$x_{\rm CH_3CN}$ & $2.5\times10^{-11}$ \\
\hline
\end{tabular}
\label{tab:chemmodel}
\end{table}

\begin{figure*}
\includegraphics[clip=,width=0.9\textwidth]{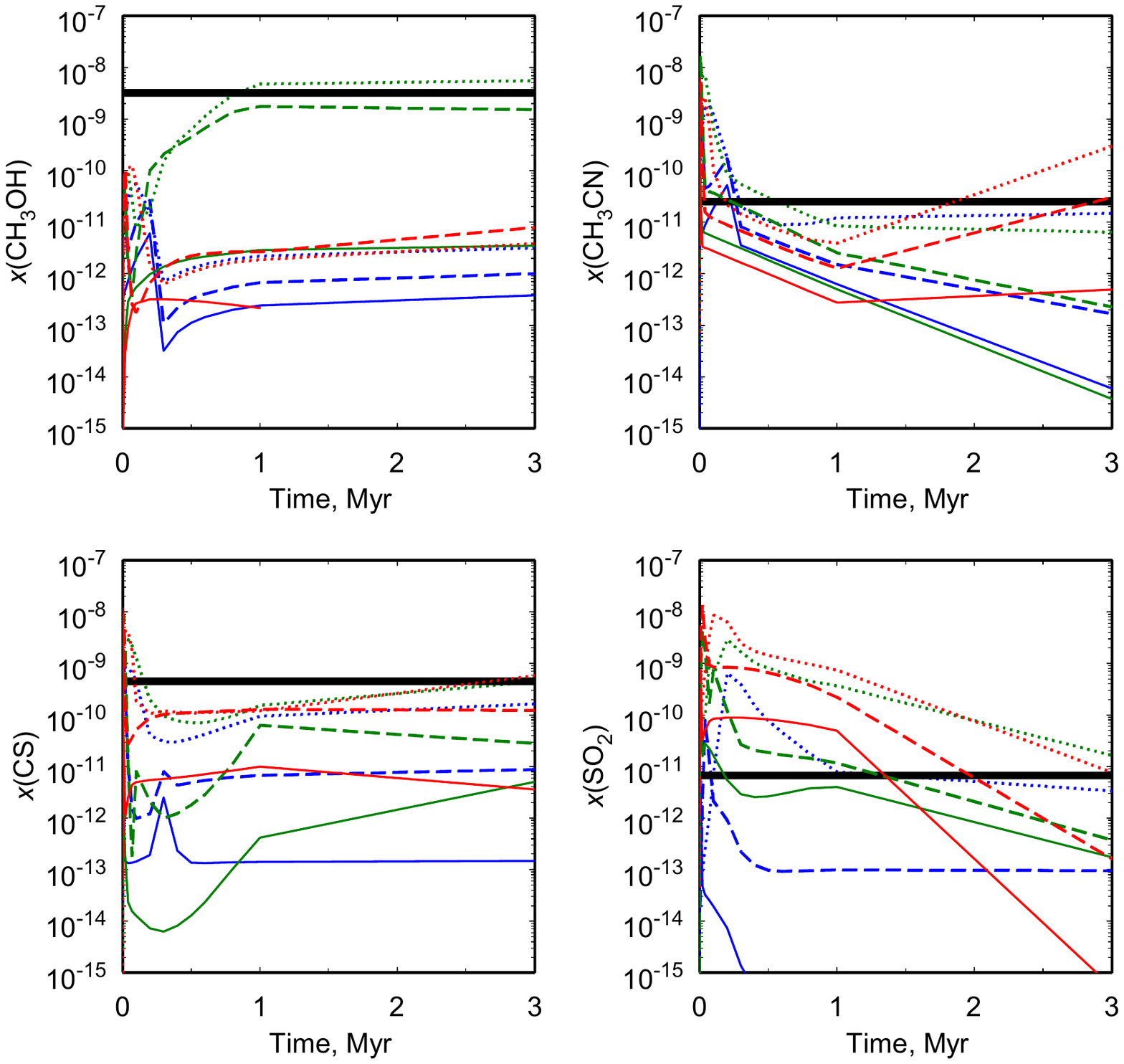}
\caption{Results of astrochemical modelling for Core~2. Colours are used to indicated gas/dust temperature, with blue denoting 10\,K, green denoting 20\,K, and red denoting 30\,K. Line styles correspond to different density values, dotted being $10^4$~cm$^{-3}$, dashed being $10^5$~cm$^{-3}$, and solid being $10^6$~cm$^{-3}$. The thick black lines indicate the abundances for each species, obtained from the observations.}
\label{chemfig1}
\end{figure*}

First, we run a small set of models for number densities $n_{\rm H}$ from $10^4$~cm$^{-3}$ to $10^6$~cm$^{-3}$ and for temperatures from 10\,K to 30\,K (gas and dust temperatures are assumed to be equal), neglecting any UV irradiation, except for cosmic-ray-induced photons. The results, shown in Fig.~\ref{chemfig1}, indicate that the observed methanol abundance can only be reproduced in models with $T=20$\,K, $n_{\rm H}=10^4-10^5$~cm$^{-3}$, and a chemical age of about 1~Myr or greater. This age should be treated with caution, as its zero moment probably predates the formation of RCW~120. The abundances of CH$_3$CN and CS give more weight to $n_{\rm H}=10^4$~cm$^{-3}$ (dotted lines in Fig.~\ref{chemfig1}), while the abundance of SO$_2$ apparently indicates $n_{\rm H}=10^5$~cm$^{-3}$ (dashed lines in Fig.~\ref{chemfig1}). These results should not be considered as mutually exclusive, as it is quite possible that the observed lines of various molecules do not arise in exactly the same conditions.

Given the high H$_2$ column density ($3.2\times 10^{23}$\,cm$^{-2}$) and the small physical size of the cores ($\sim0.1$~pc), the average density is, indeed, of the order of $10^5$~cm$^{-3}$. However, our modelling seemingly indicates that higher densities, $\ge10^6$~cm$^{-3}$, in combination with the low temperature, are excluded for methanol, as in that dense medium it mostly resides on dust grain surfaces \citep[e.g.][]{2006A&A...457..927G, 2011A&A...533A..24W, Punanova_2018}. It is possible to obtain both high density and high gas-phase methanol abundance by adding a warm-up phase. We treat this in a simple manner, by considering a second stage with the same density of $10^6$~cm$^{-3}$, but with elevated temperatures. To reproduce the observed abundance, a growth of the gas/dust temperature up to $\sim90$\,K or higher is needed to evaporate a sufficient amount of methanol from icy mantles for the adopted desorption energy of 4235\,K \citep{2017SSRv..212....1C}. It should be kept in mind that even in this case, the gas-phase methanol abundance rapidly drops due to enhanced ion-neutral chemistry. Also, after the temperature increase, the abundances of other species grow well above the observed values. This again hints that the emission regions for the various molecules are not spatially coincident. For example, there is a possibility that the methanol emission is associated with outflows (see Sec.~\ref{sec:disc}), and we find line wings on the observed spectra of methanol in contrast with the CS spectra. Note that the abundances of CH$_3$CN, CS and SO$_2$ can be underestimated because of the unknown beam filling factor for these molecules.

In summary, we conclude that our estimates for the abundances of CH$_3$OH, CH$_3$CN (low-excitation value), CS and SO$_2$ are reasonable. The specific values of the abundances hint at the warm-up phase in Core~2. The detection of the high-excitation CH$_3$CN lines suggest that the central source, responsible for the warm up, might develop a compact hot core, but does not provide enough warming to enhance the abundances up to values typical for hot gas. Therefore, we may be observing the beginning of hot-core chemistry in Core~2. To extract more clues from the molecular composition of the cores, more thorough modelling is needed to take into account the physical and evolutionary complexity of the region.

\section{Discussion}\label{sec:disc}

Core~1 and Core~2 consist of two and five fragments, respectively, on a 0.01~pc scale, according to the ALMA~3-mm continuum observations made by \citet{Figueira_2018}. The most massive fragment is found in Core~2, with an estimated mass of 27~M$_{\rm \odot}$ with a factor of 2 for the absolute mass uncertainty. \citet{Figueira_2018} detected CH$_3$CN and SO$_2$ lines there and proposed the existence of an uc\hii{} region on the basis of the detection of both of these lines. However, no radio continuum emission point sources was found toward Clump~1 in the NVSS image \citep{Kirsanova_2019}.

\citet{Voronkov_2014} detected six methanol maser spots at 36~GHz and seven spots at 44~GHz towards Clump~1. Four of these are observed in the same directions (their source G348.18+0.48). These Class~I methanol masers arise via collision pumping by molecular hydrogen \citep[e.g.][]{Cragg_1992, Leurini_2016} and are often considered as tracers of interface regions of outflows \citep[see, e.g.][]{Voronkov_2012}. Indeed, the maser spectra in the spots observed by \citet{Voronkov_2014} have profiles with significant wings, indicating complex kinematics and probably outflows in Clump~1. The wings of the thermal methanol lines reported by us suggest that both types of the methanol emission (maser and partly thermal) arise in the outflows. Water vapour masers were detected here by \citet{Braz_1982}, which are also associated with collision pumping and molecular outflows \citep[e.g.][]{Genzel_1981, Gwinn_1992, Reid_1995}, but sometimes also with uc\hii{} regions~\citep{Hunter_1994}. However, no methanol maser at 6.7~GHz was found here during the Methanol Multibeam Survey \citep{Caswell_2010}, while the 6.7~GHz masers are considered as excellent signposts of the location of massive YSOs~\citep[e.g.][]{Breen_2013} and often observed near uc\hii{} regions \citep[e.g.][]{vanderWalt_2003}. 

We show that warming-up gas-phase chemistry, related with the appearance of a heating source, is observed in Core~2. The methanol abundance in Core~2 found here is higher than observed in massive and low-mass dark clouds \citep[e.g.][]{Friberg_1988, Kalenskii_1994, Dickens_2000, Sanhueza_2013, 2014ApJ...780...85V, Punanova_2018}, but the upper limit of the methanol abundance is 10--100 times lower, or, in rare cases, consistent with the lower limits found in hot cores or uc\hii{} regions \citep[e.g.][]{Schoier_02, Sutton_2004, Garrod_2008, Purcell_2009, 2014A&A...569A..11S, 2018ApJ...853..152M, 2020arXiv201102226G}. As found by \citet{Remijan_2004, Kalenskii_2000, 2014A&A...569A..11S, 2014ApJ...786...38H, 2018ApJ...853..152M, 2020ApJ...898...54T}, the typical relative abundance of CH$_3$CN in hot cores is $10^{-9}-10^{-7}$, which is also much higher than the abundance obtained with the detected $k \leq 3$ transitions in Sec.~\ref{sec:ch3cn}. The study of \citet{Kalenskii_2000} also contains a list of high-mass YSOs with low CH$_3$CN abundances. These sources might contain hot regions accompanied by colder gas, exactly as we see in Core~2. \citet{Fayolle_2015} studied poorly explored organic-poor massive young stellar objects with weak hot organic molecular lines and found that complex organic molecules might be released from dust grains not only at the hot core stage, but earlier, depending on ice composition and temperature.

The CH$_3$OH and CH$_3$CN molecules are considered to be parent and daughter species in chemical models of gas-grain chemistry. This is why the observations of these lines show the ratio of relative abundances $[\rm CH_3CN/CH_3OH]\le1$. \citet{2018A&A...618A.145O} combined measurements of the $[\rm CH_3CN/CH_3OH]$ ratio in protostars of various masses and found values between 0.002--0.006 in low- and intermediate-mass regions. However, in high-mass regions, this ratio shows a range of values of up to two orders of magnitude: from 0.25 in G29.96 and 0.1 in Orion-KL~\citep{2009AJ....137..406B} and Sgr~B2 \citep{2013A&A...559A..47B}, to $0.01-0.04$ in several hot cores \citep{2007A&A...465..913B, 2018ApJ...853..152M}, and 0.0025--0.01 in the massive young protostellar object G345.4938+01.4677 \citep[Cores~C and B, respectively,][]{2020ApJ...898...54T}. Our value is close to the last value detected in G331~Core~B (hot core), which is more evolved than Core~C with a lower $[\rm CH_3CN/CH_3OH]$ ratio. Therefore, the observed $[\rm CH_3CN/CH_3OH]$ ratio agrees with our conclusion about the warm up and beginning of the hot-core stage in the Core~2.

Due to low $E_{\rm u}$ values (see Table~\ref{tab:detected}), \cs($5-4$) and SO$_2$($5_{2,4}-4_{1,3}$) lines are not straightforward signposts of a hot core stage. The detection of lines of such molecules as CS and SO$_2$ suggests that the cold phase of protostellar evolution has completed. During the warm up phase, H$_2$S molecules evaporate from grain mantles and give rise to gas-phase sulfur chemistry, producing SO, SO$_2$, CS, etc. \citep[see, e.g. chemical modelling by][]{Charnley_1997}. However, the abundance of SO$_2$ found in this work is up to several orders lower than the typical value found at the hot core stage as shown in the review of several hot cores by~\citet{Schoier_02} and \citet{Wakelam_2004}. The value of $x_{\rm CS} \approx 10^{-10}-10^{-9}$ obtained in this study is typical for the stage of freeze-out of molecules onto dust grains \citep[e.g.][]{Schoier_02}. \citet{2002ApJ...566..945B} studied several dozen high-mass protostellar candidates and found an average abundance of $x_{\rm CS} \approx 8\times 10^{-9}$, which is about an order of magnitude higher than our value in Core~2. More-evolved stages of uc\hii{} regions show similar abundances~\citep[e.~g.]{2009MNRAS.395.2234Z}. Therefore, the cold stage seems to have finished quite recently, and the abundances are still  quite low. We note again that our CS and SO$_2$ abundances are obtained using the same filling factor as we found from the analysis of the methanol emission. If the emission of SO$_2$ and CS arise in less extended regions as compared to methanol, the SO$_2$ and CS filling factors should be smaller; therefore, the abundances will be higher.

Bearing in mind the studies mentioned above, and our own results, we state that the Core~2 cannot be considered as an example of a high-mass analogue of a pre-stellar Class~0 object. We find warm-up chemistry and signatures of the beginning of the hot core stage there.

\section{Conclusions}\label{sec:conc}

We report the detection of methanol lines towards massive Core~1 and Core~2, belonging to the neutral envelope of the \hii{} region RCW~120, and also the detection of CH$_3$CN, CS and SO$_2$ lines towards Core~2. We estimate the gas physical parameters using the methanol lines and obtain gas temperature less than 100~K in both regions. The value of $n_{\rm H_2}$ is in the range of $10^5-10^7$~cm$^{-3}$ in Core~2, and has a higher uncertainty in Core~1. However, the detection of high-excitation CH$_3$CN lines in Core~2 indicates that the region might contain hot gas. The relative abundances of CH$_3$OH, CS, SO$_2$ and CH$_3$CN found in the present study are quite low for a hot-core stage, but agree with a model of warm chemistry in dense gas. We suggest that Core~2 is in the beginning of the hot core stage.

\section*{Acknowledgements}

We are thankful to the staff of the Onsala Space Observatory for their care for technical details of the O-083.F-9311A-2009 project, to Maxim Voronkov for discussions of the methanol maser activity and the referee for comments and suggestions which helped us to clarify our statements and improve the presentation. Astrochemical modelling performed by D. S. Wiebe was supported by the RFBR grant 20-02-00643. M. S. Kirsanova and P. A. Boley were supported by RSF, research project 18-72-10132. S. V. Salii and A. M. Sobolev work was supported by the Ministry of Education and  Science  of  the  Russian  Federation  within  the  framework  of  the  research  activities  (project  no.  FEUZ-2020-0030), as well as Decree 211 of the Government of Russian Federation (contract no. 02.A03.21.0006). This research has made use of NASA's Astrophysics Data System Bibliographic Services; SIMBAD database, operated at CDS, Strasbourg, France~\citep{Wenger_2000}, Astropy,\footnote{http://www.astropy.org} a community-developed core Python package for Astronomy \citep{astropy:2013, astropy:2018}, Matplotlib \citep{Hunter:2007} and APLpy, an open-source plotting package for Python \citet{2012ascl.soft08017R}.

\section*{Data Availability}

The data underlying this article will be shared on reasonable request to the corresponding author.

\bibliographystyle{mnras}
\bibliography{rcw120_cond1}

\begin{thebibliography}{}
\makeatletter
\relax
\def\mn@urlcharsother{\let\do\@makeother \do\$\do\&\do\#\do\^\do\_\do\%\do\~}
\def\mn@doi{\begingroup\mn@urlcharsother \@ifnextchar [ {\mn@doi@}
  {\mn@doi@[]}}
\def\mn@doi@[#1]#2{\def\@tempa{#1}\ifx\@tempa\@empty \href
  {http://dx.doi.org/#2} {doi:#2}\else \href {http://dx.doi.org/#2} {#1}\fi
  \endgroup}
\def\mn@eprint#1#2{\mn@eprint@#1:#2::\@nil}
\def\mn@eprint@arXiv#1{\href {http://arxiv.org/abs/#1} {{\tt arXiv:#1}}}
\def\mn@eprint@dblp#1{\href {http://dblp.uni-trier.de/rec/bibtex/#1.xml}
  {dblp:#1}}
\def\mn@eprint@#1:#2:#3:#4\@nil{\def\@tempa {#1}\def\@tempb {#2}\def\@tempc
  {#3}\ifx \@tempc \@empty \let \@tempc \@tempb \let \@tempb \@tempa \fi \ifx
  \@tempb \@empty \def\@tempb {arXiv}\fi \@ifundefined
  {mn@eprint@\@tempb}{\@tempb:\@tempc}{\expandafter \expandafter \csname
  mn@eprint@\@tempb\endcsname \expandafter{\@tempc}}}

\bibitem[\protect\citeauthoryear{{Anderson} et~al.,}{{Anderson}
  et~al.}{2012}]{Anderson_2012}
{Anderson} L.~D.,  et~al., 2012, \mn@doi [\aap] {10.1051/0004-6361/201117283},
  \href {http://adsabs.harvard.edu/abs/2012A%26A...542A..10A} {542, A10}

\bibitem[\protect\citeauthoryear{{Astropy Collaboration} et~al.,}{{Astropy
  Collaboration} et~al.}{2013}]{astropy:2013}
{Astropy Collaboration} et~al., 2013, \mn@doi [\aap]
  {10.1051/0004-6361/201322068}, \href
  {http://adsabs.harvard.edu/abs/2013A\%26A...558A..33A} {558, A33}

\bibitem[\protect\citeauthoryear{{Belitsky} et~al.,}{{Belitsky}
  et~al.}{2006}]{Belitsky_2006}
{Belitsky} V.,  et~al., 2006, in Society of Photo-Optical Instrumentation
  Engineers (SPIE) Conference Series. p. 62750G, \mn@doi{10.1117/12.671383}

\bibitem[\protect\citeauthoryear{{Belloche}, {M{\"u}ller}, {Menten}, {Schilke}
  \& {Comito}}{{Belloche} et~al.}{2013}]{2013A&A...559A..47B}
{Belloche} A.,  {M{\"u}ller} H.~S.~P.,  {Menten} K.~M.,  {Schilke} P.,
  {Comito} C.,  2013, \mn@doi [\aap] {10.1051/0004-6361/201321096}, \href
  {https://ui.adsabs.harvard.edu/abs/2013A&A...559A..47B} {559, A47}

\bibitem[\protect\citeauthoryear{{Beltr{\'a}n} et~al.,}{{Beltr{\'a}n}
  et~al.}{2018}]{Beltran_2018}
{Beltr{\'a}n} M.~T.,  et~al., 2018, \mn@doi [\aap]
  {10.1051/0004-6361/201832811}, \href
  {https://ui.adsabs.harvard.edu/abs/2018A&A...615A.141B} {615, A141}

\bibitem[\protect\citeauthoryear{{Beuther}, {Schilke}, {Menten}, {Motte},
  {Sridharan}  \& {Wyrowski}}{{Beuther} et~al.}{2002}]{2002ApJ...566..945B}
{Beuther} H.,  {Schilke} P.,  {Menten} K.~M.,  {Motte} F.,  {Sridharan} T.~K.,
   {Wyrowski} F.,  2002, \mn@doi [\apj] {10.1086/338334}, \href
  {https://ui.adsabs.harvard.edu/abs/2002ApJ...566..945B} {566, 945}

\bibitem[\protect\citeauthoryear{{Beuther}, {Zhang}, {Bergin}  \&
  {Sridharan}}{{Beuther} et~al.}{2009}]{2009AJ....137..406B}
{Beuther} H.,  {Zhang} Q.,  {Bergin} E.~A.,   {Sridharan} T.~K.,  2009, \mn@doi
  [\aj] {10.1088/0004-6256/137/1/406}, \href
  {https://ui.adsabs.harvard.edu/abs/2009AJ....137..406B} {137, 406}

\bibitem[\protect\citeauthoryear{{Bisschop}, {J{\o}rgensen}, {van Dishoeck}  \&
  {de Wachter}}{{Bisschop} et~al.}{2007}]{2007A&A...465..913B}
{Bisschop} S.~E.,  {J{\o}rgensen} J.~K.,  {van Dishoeck} E.~F.,   {de Wachter}
  E.~B.~M.,  2007, \mn@doi [\aap] {10.1051/0004-6361:20065963}, \href
  {https://ui.adsabs.harvard.edu/abs/2007A&A...465..913B} {465, 913}

\bibitem[\protect\citeauthoryear{{Braz} \& {Scalise}}{{Braz} \&
  {Scalise}}{1982}]{Braz_1982}
{Braz} M.~A.,  {Scalise} Jr. E.,  1982, \aap, \href
  {http://adsabs.harvard.edu/abs/1982A%26A...107..272B} {107, 272}

\bibitem[\protect\citeauthoryear{{Breen}, {Ellingsen}, {Contreras}, {Green},
  {Caswell}, {Stevens}, {Dawson}  \& {Voronkov}}{{Breen}
  et~al.}{2013}]{Breen_2013}
{Breen} S.~L.,  {Ellingsen} S.~P.,  {Contreras} Y.,  {Green} J.~A.,  {Caswell}
  J.~L.,  {Stevens} J.~B.,  {Dawson} J.~R.,   {Voronkov} M.~A.,  2013, \mn@doi
  [\mnras] {10.1093/mnras/stt1315}, \href
  {http://adsabs.harvard.edu/abs/2013MNRAS.435..524B} {435, 524}

\bibitem[\protect\citeauthoryear{{Caswell} et~al.,}{{Caswell}
  et~al.}{2010}]{Caswell_2010}
{Caswell} J.~L.,  et~al., 2010, \mn@doi [\mnras]
  {10.1111/j.1365-2966.2010.16339.x}, \href
  {http://adsabs.harvard.edu/abs/2010MNRAS.404.1029C} {404, 1029}

\bibitem[\protect\citeauthoryear{{Charnley}}{{Charnley}}{1997}]{Charnley_1997}
{Charnley} S.~B.,  1997, \mn@doi [\apj] {10.1086/304011}, \href
  {http://adsabs.harvard.edu/abs/1997ApJ...481..396C} {481, 396}

\bibitem[\protect\citeauthoryear{{Cragg}, {Johns}, {Godfrey}  \&
  {Brown}}{{Cragg} et~al.}{1992}]{Cragg_1992}
{Cragg} D.~M.,  {Johns} K.~P.,  {Godfrey} P.~D.,   {Brown} R.~D.,  1992,
  \mn@doi [\mnras] {10.1093/mnras/259.1.203}, \href
  {http://adsabs.harvard.edu/abs/1992MNRAS.259..203C} {259, 203}

\bibitem[\protect\citeauthoryear{{Cuppen}, {Walsh}, {Lamberts}, {Semenov},
  {Garrod}, {Penteado}  \& {Ioppolo}}{{Cuppen}
  et~al.}{2017}]{2017SSRv..212....1C}
{Cuppen} H.~M.,  {Walsh} C.,  {Lamberts} T.,  {Semenov} D.,  {Garrod} R.~T.,
  {Penteado} E.~M.,   {Ioppolo} S.,  2017, \mn@doi [\ssr]
  {10.1007/s11214-016-0319-3}, \href
  {https://ui.adsabs.harvard.edu/abs/2017SSRv..212....1C} {212, 1}

\bibitem[\protect\citeauthoryear{{Deharveng}, {Zavagno}, {Schuller}, {Caplan},
  {Pomar{\`e}s}  \& {De Breuck}}{{Deharveng} et~al.}{2009}]{Deharveng_09}
{Deharveng} L.,  {Zavagno} A.,  {Schuller} F.,  {Caplan} J.,  {Pomar{\`e}s} M.,
    {De Breuck} C.,  2009, \mn@doi [\aap] {10.1051/0004-6361/200811337}, \href
  {http://adsabs.harvard.edu/abs/2009A%26A...496..177D} {496, 177}

\bibitem[\protect\citeauthoryear{{Dickens}, {Irvine}, {Snell}, {Bergin},
  {Schloerb}, {Pratap}  \& {Miralles}}{{Dickens} et~al.}{2000}]{Dickens_2000}
{Dickens} J.~E.,  {Irvine} W.~M.,  {Snell} R.~L.,  {Bergin} E.~A.,  {Schloerb}
  F.~P.,  {Pratap} P.,   {Miralles} M.~P.,  2000, \mn@doi [\apj]
  {10.1086/317040}, \href {http://adsabs.harvard.edu/abs/2000ApJ...542..870D}
  {542, 870}

\bibitem[\protect\citeauthoryear{{Endres}, {Schlemmer}, {Schilke}, {Stutzki}
  \& {M{\"u}ller}}{{Endres} et~al.}{2016}]{Endres_2016}
{Endres} C.~P.,  {Schlemmer} S.,  {Schilke} P.,  {Stutzki} J.,   {M{\"u}ller}
  H.~S.~P.,  2016, \mn@doi [Journal of Molecular Spectroscopy]
  {10.1016/j.jms.2016.03.005}, \href
  {http://adsabs.harvard.edu/abs/2016JMoSp.327...95E} {327, 95}

\bibitem[\protect\citeauthoryear{{Fayolle}, {{\"O}berg}, {Garrod}, {van
  Dishoeck}  \& {Bisschop}}{{Fayolle} et~al.}{2015}]{Fayolle_2015}
{Fayolle} E.~C.,  {{\"O}berg} K.~I.,  {Garrod} R.~T.,  {van Dishoeck} E.~F.,
  {Bisschop} S.~E.,  2015, \mn@doi [\aap] {10.1051/0004-6361/201323114}, \href
  {http://adsabs.harvard.edu/abs/2015A%26A...576A..45F} {576, A45}

\bibitem[\protect\citeauthoryear{{Figueira} et~al.,}{{Figueira}
  et~al.}{2017}]{Figueira_2017}
{Figueira} M.,  et~al., 2017, \mn@doi [\aap] {10.1051/0004-6361/201629379},
  \href {http://esoads.eso.org/abs/2017A%26A...600A..93F} {600, A93}

\bibitem[\protect\citeauthoryear{{Figueira}, {Bronfman}, {Zavagno}, {Louvet},
  {Lo}, {Finger}  \& {Rod{\'o}n}}{{Figueira} et~al.}{2018}]{Figueira_2018}
{Figueira} M.,  {Bronfman} L.,  {Zavagno} A.,  {Louvet} F.,  {Lo} N.,  {Finger}
  R.,   {Rod{\'o}n} J.,  2018, \mn@doi [\aap] {10.1051/0004-6361/201832930},
  \href {http://adsabs.harvard.edu/abs/2018A%26A...616L..10F} {616, L10}

\bibitem[\protect\citeauthoryear{{Figueira}, {Zavagno}, {Bronfman}, {Russeil},
  {Finger}  \& {Schuller}}{{Figueira} et~al.}{2020}]{Figueira_2020}
{Figueira} M.,  {Zavagno} A.,  {Bronfman} L.,  {Russeil} D.,  {Finger} R.,
  {Schuller} F.,  2020, \mn@doi [\aap] {10.1051/0004-6361/202037713}, \href
  {https://ui.adsabs.harvard.edu/abs/2020A&A...639A..93F} {639, A93}

\bibitem[\protect\citeauthoryear{{Fontani}, {Giannetti}, {Beltr{\'a}n},
  {Dodson}, {Rioja}, {Brand}, {Caselli}  \& {Cesaroni}}{{Fontani}
  et~al.}{2012}]{2012MNRAS.423.2342F}
{Fontani} F.,  {Giannetti} A.,  {Beltr{\'a}n} M.~T.,  {Dodson} R.,  {Rioja} M.,
   {Brand} J.,  {Caselli} P.,   {Cesaroni} R.,  2012, \mn@doi [\mnras]
  {10.1111/j.1365-2966.2012.21043.x}, \href
  {https://ui.adsabs.harvard.edu/abs/2012MNRAS.423.2342F} {423, 2342}

\bibitem[\protect\citeauthoryear{{Friberg}, {Madden}, {Hjalmarson}  \&
  {Irvine}}{{Friberg} et~al.}{1988}]{Friberg_1988}
{Friberg} P.,  {Madden} S.~C.,  {Hjalmarson} A.,   {Irvine} W.~M.,  1988, \aap,
  \href {http://adsabs.harvard.edu/abs/1988A%26A...195..281F} {195, 281}

\bibitem[\protect\citeauthoryear{{Garrod} \& {Herbst}}{{Garrod} \&
  {Herbst}}{2006}]{2006A&A...457..927G}
{Garrod} R.~T.,  {Herbst} E.,  2006, \mn@doi [\aap]
  {10.1051/0004-6361:20065560}, \href
  {https://ui.adsabs.harvard.edu/abs/2006A&A...457..927G} {457, 927}

\bibitem[\protect\citeauthoryear{{Garrod}, {Widicus Weaver}  \&
  {Herbst}}{{Garrod} et~al.}{2008}]{Garrod_2008}
{Garrod} R.~T.,  {Widicus Weaver} S.~L.,   {Herbst} E.,  2008, \mn@doi [\apj]
  {10.1086/588035}, \href {http://adsabs.harvard.edu/abs/2008ApJ...682..283G}
  {682, 283}

\bibitem[\protect\citeauthoryear{{Genzel}, {Reid}, {Moran}  \&
  {Downes}}{{Genzel} et~al.}{1981}]{Genzel_1981}
{Genzel} R.,  {Reid} M.~J.,  {Moran} J.~M.,   {Downes} D.,  1981, \mn@doi
  [\apj] {10.1086/158764}, \href
  {http://adsabs.harvard.edu/abs/1981ApJ...244..884G} {244, 884}

\bibitem[\protect\citeauthoryear{{Goldsmith} \& {Langer}}{{Goldsmith} \&
  {Langer}}{1999}]{Goldsmith_1999}
{Goldsmith} P.~F.,  {Langer} W.~D.,  1999, \mn@doi [\apj] {10.1086/307195},
  \href {http://adsabs.harvard.edu/abs/1999ApJ...517..209G} {517, 209}

\bibitem[\protect\citeauthoryear{{Gorai}, {Das}, {Shimonishi}, {Sahu},
  {Mondal}, {Bhat}  \& {Chakrabarti}}{{Gorai}
  et~al.}{2020}]{2020arXiv201102226G}
{Gorai} P.,  {Das} A.,  {Shimonishi} T.,  {Sahu} D.,  {Mondal} S.~K.,  {Bhat}
  B.,   {Chakrabarti} S.~K.,  2020, arXiv e-prints, \href
  {https://ui.adsabs.harvard.edu/abs/2020arXiv201102226G} {p. arXiv:2011.02226}

\bibitem[\protect\citeauthoryear{{Gratier}, {Pety}, {Guzm{\'a}n}, {Gerin},
  {Goicoechea}, {Roueff}  \& {Faure}}{{Gratier} et~al.}{2013}]{Gratier_2013}
{Gratier} P.,  {Pety} J.,  {Guzm{\'a}n} V.,  {Gerin} M.,  {Goicoechea} J.~R.,
  {Roueff} E.,   {Faure} A.,  2013, \mn@doi [\aap]
  {10.1051/0004-6361/201321031}, \href
  {http://adsabs.harvard.edu/abs/2013A%26A...557A.101G} {557, A101}

\bibitem[\protect\citeauthoryear{{Gwinn}, {Moran}  \& {Reid}}{{Gwinn}
  et~al.}{1992}]{Gwinn_1992}
{Gwinn} C.~R.,  {Moran} J.~M.,   {Reid} M.~J.,  1992, \mn@doi [\apj]
  {10.1086/171493}, \href {http://adsabs.harvard.edu/abs/1992ApJ...393..149G}
  {393, 149}

\bibitem[\protect\citeauthoryear{{Hern{\'a}ndez-Hern{\'a}ndez}, {Zapata},
  {Kurtz}  \& {Garay}}{{Hern{\'a}ndez-Hern{\'a}ndez}
  et~al.}{2014}]{2014ApJ...786...38H}
{Hern{\'a}ndez-Hern{\'a}ndez} V.,  {Zapata} L.,  {Kurtz} S.,   {Garay} G.,
  2014, \mn@doi [\apj] {10.1088/0004-637X/786/1/38}, \href
  {https://ui.adsabs.harvard.edu/abs/2014ApJ...786...38H} {786, 38}

\bibitem[\protect\citeauthoryear{Hunter}{Hunter}{2007}]{Hunter:2007}
Hunter J.~D.,  2007, \mn@doi [Computing in Science \& Engineering]
  {10.1109/MCSE.2007.55}, 9, 90

\bibitem[\protect\citeauthoryear{{Hunter}, {Taylor}, {Felli}  \&
  {Tofani}}{{Hunter} et~al.}{1994}]{Hunter_1994}
{Hunter} T.~R.,  {Taylor} G.~B.,  {Felli} M.,   {Tofani} G.,  1994, \aap, \href
  {http://adsabs.harvard.edu/abs/1994A%26A...284..215H} {284, 215}

\bibitem[\protect\citeauthoryear{{Kalenskii} \& {Sobolev}}{{Kalenskii} \&
  {Sobolev}}{1994}]{Kalenskii_1994}
{Kalenskii} S.~V.,  {Sobolev} A.~M.,  1994, Astronomy Letters, \href
  {http://adsabs.harvard.edu/abs/1994AstL...20...91K} {20, 91}

\bibitem[\protect\citeauthoryear{{Kalenskii}, {Promislov}, {Alakoz}, {Winnberg}
   \& {Johansson}}{{Kalenskii} et~al.}{2000}]{Kalenskii_2000}
{Kalenskii} S.~V.,  {Promislov} V.~G.,  {Alakoz} A.~V.,  {Winnberg} A.,
  {Johansson} L.~E.~B.,  2000, \mn@doi [Astronomy Reports] {10.1134/1.1320498},
  \href {http://adsabs.harvard.edu/abs/2000ARep...44..725K} {44, 725}

\bibitem[\protect\citeauthoryear{{Kalenskii}, {Slysh}  \&
  {Val'tts}}{{Kalenskii} et~al.}{2002}]{Kalenskii_2002}
{Kalenskii} S.~V.,  {Slysh} V.~I.,   {Val'tts} I.~E.,  2002, \mn@doi [Astronomy
  Reports] {10.1134/1.1451923}, \href
  {http://adsabs.harvard.edu/abs/2002ARep...46...96K} {46, 96}

\bibitem[\protect\citeauthoryear{{Kirsanova}, {Salii}, {Sobolev}, {Olofsson},
  {Ladeyschikov}  \& {Thomasson}}{{Kirsanova} et~al.}{2017}]{Kirsanova_2017}
{Kirsanova} M.~S.,  {Salii} S.~V.,  {Sobolev} A.~M.,  {Olofsson} A.~O.~H.,
  {Ladeyschikov} D.~A.,   {Thomasson} M.,  2017, \mn@doi [Open Astronomy]
  {10.1515/astro-2017-0020}, \href
  {http://adsabs.harvard.edu/abs/2017OAst...26...99K} {26, 99}

\bibitem[\protect\citeauthoryear{{Kirsanova}, {Pavlyuchenkov}, {Wiebe},
  {Boley}, {Salii}, {Kalenskii}, {Sobolev}  \& {Anderson}}{{Kirsanova}
  et~al.}{2019}]{Kirsanova_2019}
{Kirsanova} M.~S.,  {Pavlyuchenkov} Y.~N.,  {Wiebe} D.~S.,  {Boley} P.~A.,
  {Salii} S.~V.,  {Kalenskii} S.~V.,  {Sobolev} A.~M.,   {Anderson} L.~D.,
  2019, \mn@doi [\mnras] {10.1093/mnras/stz2048}, \href
  {https://ui.adsabs.harvard.edu/abs/2019MNRAS.488.5641K} {488, 5641}

\bibitem[\protect\citeauthoryear{{Kochina}, {Wiebe}, {Kalenskii}  \&
  {Vasyunin}}{{Kochina} et~al.}{2013}]{presta}
{Kochina} O.~V.,  {Wiebe} D.~S.,  {Kalenskii} S.~V.,   {Vasyunin} A.~I.,  2013,
  \mn@doi [Astronomy Reports] {10.1134/S1063772913110036}, \href
  {https://ui.adsabs.harvard.edu/abs/2013ARep...57..818K} {57, 818}

\bibitem[\protect\citeauthoryear{{Leurini}, {Schilke}, {Menten}, {Flower},
  {Pottage}  \& {Xu}}{{Leurini} et~al.}{2004}]{Leurini_2004}
{Leurini} S.,  {Schilke} P.,  {Menten} K.~M.,  {Flower} D.~R.,  {Pottage}
  J.~T.,   {Xu} L.-H.,  2004, \mn@doi [\aap] {10.1051/0004-6361:20047046},
  \href {http://adsabs.harvard.edu/abs/2004A%26A...422..573L} {422, 573}

\bibitem[\protect\citeauthoryear{{Leurini}, {Schilke}, {Wyrowski}  \&
  {Menten}}{{Leurini} et~al.}{2007}]{Leurini_2007}
{Leurini} S.,  {Schilke} P.,  {Wyrowski} F.,   {Menten} K.~M.,  2007, \mn@doi
  [\aap] {10.1051/0004-6361:20054245}, \href
  {http://adsabs.harvard.edu/abs/2007A%26A...466..215L} {466, 215}

\bibitem[\protect\citeauthoryear{{Leurini}, {Menten}  \& {Walmsley}}{{Leurini}
  et~al.}{2016}]{Leurini_2016}
{Leurini} S.,  {Menten} K.~M.,   {Walmsley} C.~M.,  2016, \mn@doi [\aap]
  {10.1051/0004-6361/201527974}, \href
  {http://adsabs.harvard.edu/abs/2016A%26A...592A..31L} {592, A31}

\bibitem[\protect\citeauthoryear{{Liu}, {Su}, {Zinchenko}, {Wang}, {Meyer},
  {Wang}  \& {Hsieh}}{{Liu} et~al.}{2020}]{2020ApJ...904..181L}
{Liu} S.-Y.,  {Su} Y.-N.,  {Zinchenko} I.,  {Wang} K.-S.,  {Meyer} D. M.~A.,
  {Wang} Y.,   {Hsieh} I.~T.,  2020, \mn@doi [\apj] {10.3847/1538-4357/abc0ec},
  \href {https://ui.adsabs.harvard.edu/abs/2020ApJ...904..181L} {904, 181}

\bibitem[\protect\citeauthoryear{{Mangum} \& {Shirley}}{{Mangum} \&
  {Shirley}}{2015}]{Mangum_2015}
{Mangum} J.~G.,  {Shirley} Y.~L.,  2015, \mn@doi [\pasp] {10.1086/680323},
  \href {http://adsabs.harvard.edu/abs/2015PASP..127..266M} {127, 266}

\bibitem[\protect\citeauthoryear{{Marsh}, {Whitworth}  \& {Lomax}}{{Marsh}
  et~al.}{2015}]{2015MNRAS.454.4282M}
{Marsh} K.~A.,  {Whitworth} A.~P.,   {Lomax} O.,  2015, \mn@doi [\mnras]
  {10.1093/mnras/stv2248}, \href
  {https://ui.adsabs.harvard.edu/abs/2015MNRAS.454.4282M} {454, 4282}

\bibitem[\protect\citeauthoryear{{Marsh} et~al.,}{{Marsh}
  et~al.}{2017}]{2017MNRAS.471.2730M}
{Marsh} K.~A.,  et~al., 2017, \mn@doi [\mnras] {10.1093/mnras/stx1723}, \href
  {https://ui.adsabs.harvard.edu/abs/2017MNRAS.471.2730M} {471, 2730}

\bibitem[\protect\citeauthoryear{{Mendoza} et~al.,}{{Mendoza}
  et~al.}{2018}]{2018ApJ...853..152M}
{Mendoza} E.,  et~al., 2018, \mn@doi [\apj] {10.3847/1538-4357/aaa1ec}, \href
  {https://ui.adsabs.harvard.edu/abs/2018ApJ...853..152M} {853, 152}

\bibitem[\protect\citeauthoryear{{Minier} et~al.,}{{Minier}
  et~al.}{2013}]{Minier_2013}
{Minier} V.,  et~al., 2013, \mn@doi [\aap] {10.1051/0004-6361/201219423}, \href
  {http://esoads.eso.org/abs/2013A%26A...550A..50M} {550, A50}

\bibitem[\protect\citeauthoryear{{Olmi}, {Cesaroni}, {Neri}  \&
  {Walmsley}}{{Olmi} et~al.}{1996}]{1996A&A...315..565O}
{Olmi} L.,  {Cesaroni} R.,  {Neri} R.,   {Walmsley} C.~M.,  1996, \aap, \href
  {https://ui.adsabs.harvard.edu/abs/1996A&A...315..565O} {315, 565}

\bibitem[\protect\citeauthoryear{{Olmi}, {Araya}, {Chapin}, {Gibb}, {Hofner},
  {Martin}  \& {Poventud}}{{Olmi} et~al.}{2010}]{2010ApJ...715.1132O}
{Olmi} L.,  {Araya} E.~D.,  {Chapin} E.~L.,  {Gibb} A.,  {Hofner} P.,  {Martin}
  P.~G.,   {Poventud} C.~M.,  2010, \mn@doi [\apj]
  {10.1088/0004-637X/715/2/1132}, \href
  {https://ui.adsabs.harvard.edu/abs/2010ApJ...715.1132O} {715, 1132}

\bibitem[\protect\citeauthoryear{{Ospina-Zamudio}, {Lefloch}, {Ceccarelli},
  {Kahane}, {Favre}, {L{\'o}pez-Sepulcre}  \& {Montarges}}{{Ospina-Zamudio}
  et~al.}{2018}]{2018A&A...618A.145O}
{Ospina-Zamudio} J.,  {Lefloch} B.,  {Ceccarelli} C.,  {Kahane} C.,  {Favre}
  C.,  {L{\'o}pez-Sepulcre} A.,   {Montarges} M.,  2018, \mn@doi [\aap]
  {10.1051/0004-6361/201832857}, \href
  {https://ui.adsabs.harvard.edu/abs/2018A&A...618A.145O} {618, A145}

\bibitem[\protect\citeauthoryear{{Pavlyuchenkov}, {Wiebe}, {Fateeva}  \&
  {Vasyunina}}{{Pavlyuchenkov} et~al.}{2011}]{Pavlyuchenkov_2011}
{Pavlyuchenkov} Y.~N.,  {Wiebe} D.~S.,  {Fateeva} A.~M.,   {Vasyunina} T.~S.,
  2011, \mn@doi [Astronomy Reports] {10.1134/S1063772911010057}, \href
  {http://adsabs.harvard.edu/abs/2011ARep...55....1P} {55, 1}

\bibitem[\protect\citeauthoryear{{Price-Whelan} et~al.,}{{Price-Whelan}
  et~al.}{2018}]{astropy:2018}
{Price-Whelan} A.~M.,  et~al., 2018, \mn@doi [\aj] {10.3847/1538-3881/aabc4f},
  \href {https://ui.adsabs.harvard.edu/#abs/2018AJ....156..123T} {156, 123}

\bibitem[\protect\citeauthoryear{{Punanova} et~al.,}{{Punanova}
  et~al.}{2018}]{Punanova_2018}
{Punanova} A.,  et~al., 2018, \mn@doi [\apj] {10.3847/1538-4357/aaad09}, \href
  {http://adsabs.harvard.edu/abs/2018ApJ...855..112P} {855, 112}

\bibitem[\protect\citeauthoryear{{Purcell}, {Longmore}, {Burton}, {Walsh},
  {Minier}, {Cunningham}  \& {Balasubramanyam}}{{Purcell}
  et~al.}{2009}]{Purcell_2009}
{Purcell} C.~R.,  {Longmore} S.~N.,  {Burton} M.~G.,  {Walsh} A.~J.,  {Minier}
  V.,  {Cunningham} M.~R.,   {Balasubramanyam} R.,  2009, \mn@doi [\mnras]
  {10.1111/j.1365-2966.2008.14283.x}, \href
  {http://adsabs.harvard.edu/abs/2009MNRAS.394..323P} {394, 323}

\bibitem[\protect\citeauthoryear{{Reid}, {Argon}, {Masson}, {Menten}  \&
  {Moran}}{{Reid} et~al.}{1995}]{Reid_1995}
{Reid} M.~J.,  {Argon} A.~L.,  {Masson} C.~R.,  {Menten} K.~M.,   {Moran}
  J.~M.,  1995, \mn@doi [\apj] {10.1086/175518}, \href
  {http://adsabs.harvard.edu/abs/1995ApJ...443..238R} {443, 238}

\bibitem[\protect\citeauthoryear{{Remijan}, {Sutton}, {Snyder}, {Friedel},
  {Liu}  \& {Pei}}{{Remijan} et~al.}{2004}]{Remijan_2004}
{Remijan} A.,  {Sutton} E.~C.,  {Snyder} L.~E.,  {Friedel} D.~N.,  {Liu} S.-Y.,
    {Pei} C.-C.,  2004, \mn@doi [\apj] {10.1086/383120}, \href
  {http://adsabs.harvard.edu/abs/2004ApJ...606..917R} {606, 917}

\bibitem[\protect\citeauthoryear{{Robitaille} \& {Bressert}}{{Robitaille} \&
  {Bressert}}{2012}]{2012ascl.soft08017R}
{Robitaille} T.,  {Bressert} E.,  2012, {APLpy: Astronomical Plotting Library
  in Python} (\mn@eprint {ascl} {1208.017})

\bibitem[\protect\citeauthoryear{{Russeil}}{{Russeil}}{2003}]{Russeil_03}
{Russeil} D.,  2003, \mn@doi [\aap] {10.1051/0004-6361:20021504}, \href
  {http://adsabs.harvard.edu/abs/2003A%26A...397..133R} {397, 133}

\bibitem[\protect\citeauthoryear{{Salii}}{{Salii}}{2006}]{Salii_2006}
{Salii} S.~V.,  2006, in {Wiebe} D.~S.,  {Kirsanova} M.~S.,  eds, Star
  Formation in the Galaxy and Beyond.

\bibitem[\protect\citeauthoryear{{Salii}, {Parfenov}  \& {Sobolev}}{{Salii}
  et~al.}{2018}]{Salii_2018}
{Salii} S.,  {Parfenov} S.,   {Sobolev} A.,  2018, in Modern Star Astronomy.
  Vol. 1, Astronomy-2018 (XIII Congress of the International Public
  Organization ``Astronomical Society''). Conference Abstracts, Moscow:
  IZMIRAN, 2018. p. 276-279. pp 276--279, \mn@doi{10.31361/eaas.2018-1.062}

\bibitem[\protect\citeauthoryear{{S{\'a}nchez-Monge}
  et~al.,}{{S{\'a}nchez-Monge} et~al.}{2014}]{2014A&A...569A..11S}
{S{\'a}nchez-Monge} {\'A}.,  et~al., 2014, \mn@doi [\aap]
  {10.1051/0004-6361/201424032}, \href
  {https://ui.adsabs.harvard.edu/abs/2014A&A...569A..11S} {569, A11}

\bibitem[\protect\citeauthoryear{{Sanhueza}, {Jackson}, {Foster},
  {Jimenez-Serra}, {Dirienzo}  \& {Pillai}}{{Sanhueza}
  et~al.}{2013}]{Sanhueza_2013}
{Sanhueza} P.,  {Jackson} J.~M.,  {Foster} J.~B.,  {Jimenez-Serra} I.,
  {Dirienzo} W.~J.,   {Pillai} T.,  2013, \mn@doi [\apj]
  {10.1088/0004-637X/773/2/123}, \href
  {http://adsabs.harvard.edu/abs/2013ApJ...773..123S} {773, 123}

\bibitem[\protect\citeauthoryear{{Sch{\"o}ier}, {J{\o}rgensen}, {van Dishoeck}
  \& {Blake}}{{Sch{\"o}ier} et~al.}{2002}]{Schoier_02}
{Sch{\"o}ier} F.~L.,  {J{\o}rgensen} J.~K.,  {van Dishoeck} E.~F.,   {Blake}
  G.~A.,  2002, \mn@doi [\aap] {10.1051/0004-6361:20020756}, \href
  {http://adsabs.harvard.edu/abs/2002A%26A...390.1001S} {390, 1001}

\bibitem[\protect\citeauthoryear{{Semenov} \& {Wiebe}}{{Semenov} \&
  {Wiebe}}{2011}]{alchemic}
{Semenov} D.,  {Wiebe} D.,  2011, \mn@doi [\apjs] {10.1088/0067-0049/196/2/25},
  \href {https://ui.adsabs.harvard.edu/abs/2011ApJS..196...25S} {196, 25}

\bibitem[\protect\citeauthoryear{{Sridharan}, {Beuther}, {Saito}, {Wyrowski}
  \& {Schilke}}{{Sridharan} et~al.}{2005}]{2005ApJ...634L..57S}
{Sridharan} T.~K.,  {Beuther} H.,  {Saito} M.,  {Wyrowski} F.,   {Schilke} P.,
  2005, \mn@doi [\apjl] {10.1086/498644}, \href
  {https://ui.adsabs.harvard.edu/abs/2005ApJ...634L..57S} {634, L57}

\bibitem[\protect\citeauthoryear{{Sutton}, {Sobolev}, {Salii}, {Malyshev},
  {Ostrovskii}  \& {Zinchenko}}{{Sutton} et~al.}{2004}]{Sutton_2004}
{Sutton} E.~C.,  {Sobolev} A.~M.,  {Salii} S.~V.,  {Malyshev} A.~V.,
  {Ostrovskii} A.~B.,   {Zinchenko} I.~I.,  2004, \mn@doi [\apj]
  {10.1086/420962}, \href {http://adsabs.harvard.edu/abs/2004ApJ...609..231S}
  {609, 231}

\bibitem[\protect\citeauthoryear{{Svoboda} et~al.,}{{Svoboda}
  et~al.}{2019}]{2019ApJ...886...36S}
{Svoboda} B.~E.,  et~al., 2019, \mn@doi [\apj] {10.3847/1538-4357/ab40ca}, 886,
  36

\bibitem[\protect\citeauthoryear{{Tafalla}, {Mardones}, {Myers}, {Caselli},
  {Bachiller}  \& {Benson}}{{Tafalla} et~al.}{1998}]{Tafalla_1998}
{Tafalla} M.,  {Mardones} D.,  {Myers} P.~C.,  {Caselli} P.,  {Bachiller} R.,
  {Benson} P.~J.,  1998, \mn@doi [\apj] {10.1086/306115}, \href
  {http://adsabs.harvard.edu/abs/1998ApJ...504..900T} {504, 900}

\bibitem[\protect\citeauthoryear{{Taniguchi}, {Guzm{\'a}n}, {Majumdar}, {Saito}
   \& {Tokuda}}{{Taniguchi} et~al.}{2020}]{2020ApJ...898...54T}
{Taniguchi} K.,  {Guzm{\'a}n} A.~E.,  {Majumdar} L.,  {Saito} M.,   {Tokuda}
  K.,  2020, \mn@doi [\apj] {10.3847/1538-4357/ab994d}, \href
  {https://ui.adsabs.harvard.edu/abs/2020ApJ...898...54T} {898, 54}

\bibitem[\protect\citeauthoryear{{Tremblin} et~al.,}{{Tremblin}
  et~al.}{2013}]{Tremblin_13}
{Tremblin} P.,  et~al., 2013, \mn@doi [\aap] {10.1051/0004-6361/201322233},
  \href {http://esoads.eso.org/abs/2013A%26A...560A..19T} {560, A19}

\bibitem[\protect\citeauthoryear{{Vassilev} et~al.,}{{Vassilev}
  et~al.}{2008}]{Vassilev_2008}
{Vassilev} V.,  et~al., 2008, \mn@doi [\aap] {10.1051/0004-6361:200810459},
  \href {http://adsabs.harvard.edu/abs/2008A%26A...490.1157V} {490, 1157}

\bibitem[\protect\citeauthoryear{{Vasyunina}, {Vasyunin}, {Herbst}, {Linz},
  {Voronkov}, {Britton}, {Zinchenko}  \& {Schuller}}{{Vasyunina}
  et~al.}{2014}]{2014ApJ...780...85V}
{Vasyunina} T.,  {Vasyunin} A.~I.,  {Herbst} E.,  {Linz} H.,  {Voronkov} M.,
  {Britton} T.,  {Zinchenko} I.,   {Schuller} F.,  2014, \mn@doi [\apj]
  {10.1088/0004-637X/780/1/85}, \href
  {https://ui.adsabs.harvard.edu/abs/2014ApJ...780...85V} {780, 85}

\bibitem[\protect\citeauthoryear{{Voronkov}, {Caswell}, {Ellingsen}, {Breen},
  {Britton}, {Green}, {Sobolev}  \& {Walsh}}{{Voronkov}
  et~al.}{2012}]{Voronkov_2012}
{Voronkov} M.~A.,  {Caswell} J.~L.,  {Ellingsen} S.~P.,  {Breen} S.~L.,
  {Britton} T.~R.,  {Green} J.~A.,  {Sobolev} A.~M.,   {Walsh} A.~J.,  2012, in
  {Booth} R.~S.,  {Vlemmings} W.~H.~T.,   {Humphreys} E.~M.~L.,  eds,  IAU
  Symposium Vol. 287, Cosmic Masers - from OH to H0. pp 433--440 (\mn@eprint
  {arXiv} {1203.5492}), \mn@doi{10.1017/S174392131200748X}

\bibitem[\protect\citeauthoryear{{Voronkov}, {Caswell}, {Ellingsen}, {Green}
  \& {Breen}}{{Voronkov} et~al.}{2014}]{Voronkov_2014}
{Voronkov} M.~A.,  {Caswell} J.~L.,  {Ellingsen} S.~P.,  {Green} J.~A.,
  {Breen} S.~L.,  2014, \mn@doi [\mnras] {10.1093/mnras/stu116}, \href
  {http://adsabs.harvard.edu/abs/2014MNRAS.439.2584V} {439, 2584}

\bibitem[\protect\citeauthoryear{{Wakelam}, {Castets}, {Ceccarelli}, {Lefloch},
  {Caux}  \& {Pagani}}{{Wakelam} et~al.}{2004}]{Wakelam_2004}
{Wakelam} V.,  {Castets} A.,  {Ceccarelli} C.,  {Lefloch} B.,  {Caux} E.,
  {Pagani} L.,  2004, \mn@doi [\aap] {10.1051/0004-6361:20031572}, \href
  {http://adsabs.harvard.edu/abs/2004A%26A...413..609W} {413, 609}

\bibitem[\protect\citeauthoryear{{Ward}, {Zmuidzinas}, {Harris}  \&
  {Isaak}}{{Ward} et~al.}{2003}]{Ward2003}
{Ward} J.~S.,  {Zmuidzinas} J.,  {Harris} A.~I.,   {Isaak} K.~G.,  2003,
  \mn@doi [\apj] {10.1086/368175}, \href
  {http://adsabs.harvard.edu/abs/2003ApJ...587..171W} {587, 171}

\bibitem[\protect\citeauthoryear{{Wenger} et~al.,}{{Wenger}
  et~al.}{2000}]{Wenger_2000}
{Wenger} M.,  et~al., 2000, \mn@doi [\aaps] {10.1051/aas:2000332}, \href
  {http://adsabs.harvard.edu/abs/2000A%26AS..143....9W} {143, 9}

\bibitem[\protect\citeauthoryear{{Wiebe}, {Molyarova}, {Akimkin}, {Vorobyov}
  \& {Semenov}}{{Wiebe} et~al.}{2019}]{wiebe2019}
{Wiebe} D.~S.,  {Molyarova} T.~S.,  {Akimkin} V.~V.,  {Vorobyov} E.~I.,
  {Semenov} D.~A.,  2019, \mn@doi [\mnras] {10.1093/mnras/stz512}, \href
  {https://ui.adsabs.harvard.edu/abs/2019MNRAS.485.1843W} {485, 1843}

\bibitem[\protect\citeauthoryear{{Wilson}}{{Wilson}}{1999}]{Wilson_1999}
{Wilson} T.~L.,  1999, \mn@doi [Reports on Progress in Physics]
  {10.1088/0034-4885/62/2/002}, \href
  {http://adsabs.harvard.edu/abs/1999RPPh...62..143W} {62, 143}

\bibitem[\protect\citeauthoryear{{Wirstr{\"o}m} et~al.,}{{Wirstr{\"o}m}
  et~al.}{2011}]{2011A&A...533A..24W}
{Wirstr{\"o}m} E.~S.,  et~al., 2011, \mn@doi [\aap]
  {10.1051/0004-6361/201116525}, \href
  {https://ui.adsabs.harvard.edu/abs/2011A&A...533A..24W} {533, A24}

\bibitem[\protect\citeauthoryear{{Zavagno}, {Pomar{\`e}s}, {Deharveng},
  {Hosokawa}, {Russeil}  \& {Caplan}}{{Zavagno} et~al.}{2007}]{Zavagno_2007}
{Zavagno} A.,  {Pomar{\`e}s} M.,  {Deharveng} L.,  {Hosokawa} T.,  {Russeil}
  D.,   {Caplan} J.,  2007, \mn@doi [\aap] {10.1051/0004-6361:20077474}, \href
  {http://adsabs.harvard.edu/abs/2007A%26A...472..835Z} {472, 835}

\bibitem[\protect\citeauthoryear{{Zavagno} et~al.,}{{Zavagno}
  et~al.}{2010}]{Zavagno_2010}
{Zavagno} A.,  et~al., 2010, \mn@doi [\aap] {10.1051/0004-6361/201014623},
  \href {http://adsabs.harvard.edu/abs/2010A%26A...518L..81Z} {518, L81}

\bibitem[\protect\citeauthoryear{{Zavagno} et~al.,}{{Zavagno}
  et~al.}{2020}]{Zavagno_2020}
{Zavagno} A.,  et~al., 2020, \mn@doi [\aap] {10.1051/0004-6361/202037815},
  \href {https://ui.adsabs.harvard.edu/abs/2020A&A...638A...7Z} {638, A7}

\bibitem[\protect\citeauthoryear{{Zhang} et~al.,}{{Zhang}
  et~al.}{2020}]{2020arXiv201207738Z}
{Zhang} S.,  et~al., 2020, arXiv e-prints, \href
  {https://ui.adsabs.harvard.edu/abs/2020arXiv201207738Z} {p. arXiv:2012.07738}

\bibitem[\protect\citeauthoryear{{Zinchenko}, {Caselli}  \&
  {Pirogov}}{{Zinchenko} et~al.}{2009}]{2009MNRAS.395.2234Z}
{Zinchenko} I.,  {Caselli} P.,   {Pirogov} L.,  2009, \mn@doi [\mnras]
  {10.1111/j.1365-2966.2009.14687.x}, \href
  {https://ui.adsabs.harvard.edu/abs/2009MNRAS.395.2234Z} {395, 2234}

\bibitem[\protect\citeauthoryear{{van der Walt}, {Churchwell}, {Gaylard}  \&
  {Goedhart}}{{van der Walt} et~al.}{2003}]{vanderWalt_2003}
{van der Walt} D.~J.,  {Churchwell} E.,  {Gaylard} M.~J.,   {Goedhart} S.,
  2003, \mn@doi [\mnras] {10.1046/j.1365-8711.2003.06416.x}, \href
  {http://adsabs.harvard.edu/abs/2003MNRAS.341..270V} {341, 270}

\makeatother
\end{thebibliography}

\appendix

\section{Observed positions}

%\begin{table}
%    \centering
%    \begin{tabular}{ccc}
%    \hline
%     -30\arcsec,60\arcsec & -30\arcsec,45\arcsec & 0\arcsec,30\arcsec \\
%-30\arcsec,30\arcsec&  -60\arcsec,30\arcsec & 15\arcsec,15\arcsec \\
%-15\arcsec,15\arcsec &  -45\arcsec,15\arcsec& 30\arcsec,0\arcsec \\
%0\arcsec,0\arcsec  & -30\arcsec,0\arcsec & -60\arcsec,0\arcsec\\
%15\arcsec,-15\arcsec & -15\arcsec,-15\arcsec & 30\arcsec,-30\arcsec \\
%0\arcsec,-30\arcsec & -30\arcsec,-30\arcsec & -60\arcsec,-30\arcsec \\
%45\arcsec,-45\arcsec & 15\arcsec,-45\arcsec & 30\arcsec,-60\arcsec \\
%0\arcsec,-60\arcsec & -30\arcsec,-60\arcsec & 30\arcsec,-90\arcsec \\
%0\arcsec,-90\arcsec & & \\
%\hline
%    \end{tabular}
%    \caption{Offsets of positions where the methanol lines were observed. Reference position is $\alpha_{2000}=17^{\rm h} 12^{\rm m} 08^{\rm s}$ and $\delta_{2000}=-38^\circ$30\arcmin45\arcsec.}
%    \label{tab:observedpositions}
%\end{table}

\begin{table*}
    \centering
    \begin{tabular}{cccccccc}
    \hline
                    &                    &                     &                     &                      & --30\arcsec,60\arcsec & &\\
                    &                    &                     &                     &                      & --30\arcsec,45\arcsec & &\\
                    &                    &                     &  0\arcsec,30\arcsec &                      & --30\arcsec,30\arcsec &                      & --60\arcsec,30\arcsec\\
                    &                    & 15\arcsec,15\arcsec &                     & --15\arcsec,15\arcsec &                      & --45\arcsec,15\arcsec & \\
                    &30\arcsec,0\arcsec  &                     & 0\arcsec,0\arcsec   &                      & --30\arcsec,0\arcsec  &                      & --60\arcsec,0\arcsec  \\
                    &                    & 15\arcsec,-15\arcsec&                     &--15\arcsec,-15\arcsec &                      & & \\
                    &30\arcsec,--30\arcsec&                     & 0\arcsec,-30\arcsec &                      & --30\arcsec,--30\arcsec&                      &--60\arcsec,-30\arcsec \\
45\arcsec,-45\arcsec&                    & 15\arcsec,-45\arcsec& & & & \\
                    &30\arcsec,--60\arcsec&                     & 0\arcsec,--60\arcsec &                      &--30\arcsec,--60\arcsec&                       & \\
                    &30\arcsec,--90\arcsec&                     & 0\arcsec,--90\arcsec & & & \\
\hline
    \end{tabular}
    \caption{Offsets of positions where the methanol lines were observed. Reference position is $\alpha_{2000}=17^{\rm h} 12^{\rm m} 08^{\rm s}$ and $\delta_{2000}=-38^\circ$30\arcmin45\arcsec.}
    \label{tab:observedpositions}
\end{table*}

\bsp
\label{lastpage}
\end{document}